\documentclass[version1996]{siggraph}
\usepackage{epsf}

\def\s{\sigma}
\def\sgn{\mathop{\rm sign}\nolimits}

\def\df{\partial}
\def\nn{\nonumber}
\def\L{{\cal L}}
\def\H{{\cal H}}
\def\ph{\varphi}
\def\mod{\mathop{\mbox{mod}}\nolimits}
\def\sign{\mathop{\mbox{sign}}\nolimits}
\def\nparallel{{\not\;\!\parallel\;}}

\newlength{\mywidth}\mywidth=2.3truein 
\epsfysize=\mywidth
\epsfxsize=\mywidth
\def\fref#1{fig.\ref{#1}}

\renewenvironment{figure}{\refstepcounter{figure}
\baselineskip=0.4\normalbaselineskip\footnotesize}
{\baselineskip=\normalbaselineskip}
\def\fignum{{\bf Fig.\arabic{figure}.\quad}}

\begin{document}
\title{Structure of singularities on the world sheets 
of relativistic strings}
\author{Stanislav Klimenko*, Igor Nikitin**, Lialia Nikitina**\\
\\
{\it * Institute for High Energy Physics, Protvino, Russia}\\
{\it ** National Research Center for Information Technology, 
St.Augustin, Germany}
}
\date{}
\maketitle

\quad

\vspace{-10mm}
\begin{center}
{\bf Abstract}
\end{center}

\baselineskip=0.4\normalbaselineskip\footnotesize

We show that relativistic strings of open and closed types in Minkowski 
space-time of dimension 3 and 4 have topologically stable singular points. 
This paper describes the structure of singularities,
derives their normal forms, and introduces a local characteristic 
of singularity (topological charge), possessing 
a global law of conservation. Two other types of solutions
(breaking and exotic strings) are also considered,
which have singularities at arbitrary value of dimension.

\vspace{1mm}

\baselineskip=\normalbaselineskip\normalsize

%

\section{Introduction}

Classical string theory considers time-like surfaces of extreme area 
in $d$-dimensional Minkowski space-time, called 
{\it the world sheets of strings}. These surfaces are 
spanned in motion through the space-time of 1-dimensional object, 
called relativistic string.
The string theory is used in high energy physics to model
the inner structure of elementary particles. For this purpose 
the world sheets (WS), having microscopic spatial sizes 
and infinitely extended in temporal direction, 
are considered as structured world lines of elementary particles. 
As a result, the internal characteristics of particles, 
such as mass and spin, are expressed in terms of string dynamics 
and can be derived from a small set of fundamental constants. 
The construction of string theory encountered several hard problems, 
concerning to quantum mechanical representation of infinite-dimensional 
groups of symmetries, which nowadays are solved completely only 
in high-dimensional space-time ($d=26$) and in certain 
topologically non-trivial space-time manifolds \cite{Brink,Witten}.
Also, certain subsets in the phase space of string theory were found in 
[3-5], 
which admit anomaly-free quantization 
at arbitrary dimension of the space-time. Successful 
construction of quantum string theory is usually considered
as a first step necessary for quantization of gravitation \cite{Brink}.
On the other hand, in several works 
[6-11] 
authors noticed singular properties of the classical string theory 
in low-dimensional ($d=3,4$) flat Minkowski space-time.
In this paper we will discuss this topic in more detail. 

Let WS is represented parametrically as $x_{\mu}(\s_{1},\s_{2})$. 
Time-likeness means that the WS admits a parametrization
with $(\df_{1}x)^{2}\geq0$, $(\df_{2}x)^{2}\leq0$,
i.e. one tangent vector should be time-like or light-like,
while another one should be space-like or light-like.
The area of WS in this case can be written as

$$A=\int\!\int d\s_{1}d\s_{2}\sqrt{(\df_{1}x\;\df_{2}x)^{2}
-(\df_{1}x)^{2}(\df_{2}x)^{2}}= \mbox{extremum.}$$

\noindent{\bf Theorem 1} (type of extremum):
a regular point of WS is a saddle point of area functional.
(Proof of the theorem is placed in Appendix.)

\begin{figure}\label{f1}
\begin{center}
~\epsfysize=1.5cm\epsfxsize=4cm\epsffile{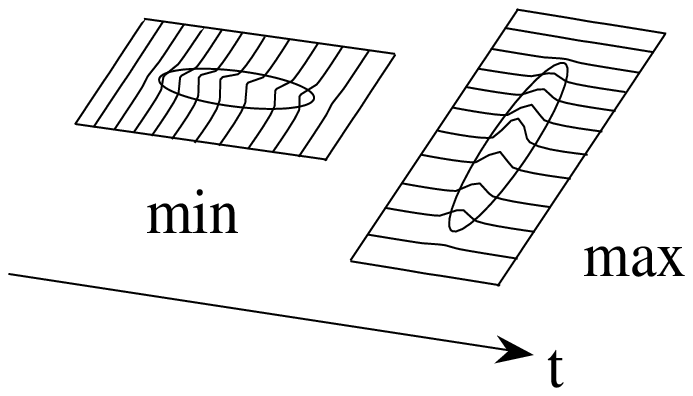}

\vspace{1mm}
\fignum Type of extremum is saddle point.
\end{center}
\end{figure}

\vspace{2mm}
The area of WS is minimal with respect to local variations,
extended in temporal direction, changing mostly the length 
of strings in equal-time slices, see \fref{f1}; and maximal for variations, 
extended in spatial direction, when mostly 
the interval of world lines of points on the string is changed.

\begin{figure}\label{f2}
\begin{center}
~\epsfysize=1.5cm\epsfxsize=6cm\epsffile{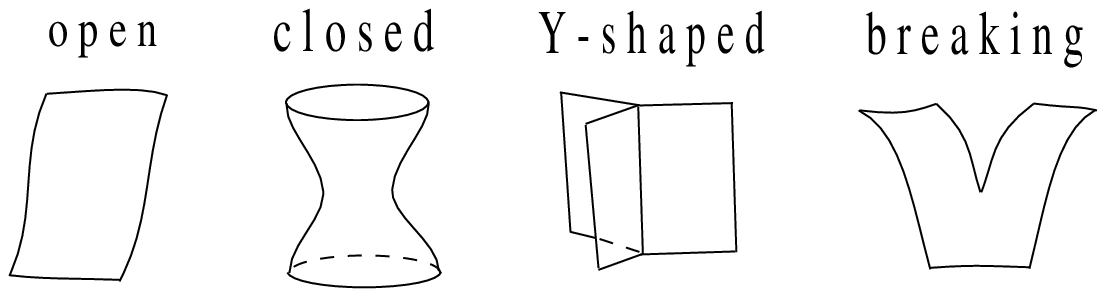}

\vspace{1mm}
\fignum Main topological types of WS.
\end{center}
\end{figure}

String theory considers WS of various topological types, see \fref{f2}:
open strings -- surfaces, homeomorphic to bands $I\times{\bf R}^{1}$, 
closed strings -- cylinders $S^{1}\times{\bf R}^{1}$,
Y-shaped strings --  3 bands, glued together along one edge,
and also the surfaces of more complex topology, corresponding to transitions
between the described types (decays and transmutations of particles).

Extremum of area for each topological type
leads to Lagrange-Euler equations, satisfied in internal points of WS, 
which have a form of local conservation of energy-momentum:
$$p_{i}^{\mu}={\delta A/\delta(\df_{i}x_{\mu})},\quad
\df_{i}p_{i}^{\mu}=0,$$ and boundary conditions, 
implying that total flow of momentum through the boundary vanishes.
For example, open string satisfies the equation 
$p_{i}^{\mu}\epsilon_{ij}d\s_{j}=0$ 
on the boundary (here $d\s_{j}$ is tangent element of WS boundary 
on the parameters plane, $\epsilon_{ij}d\s_{j}$ is normal element); 
for Y-shaped string $\sum p_{i}^{\mu}\epsilon_{ij}d\s_{j}=0$
on the world line of the node, where the sum is taken
for three surfaces attached to this line. 

Construction of string theory usually proceeds in Hamiltonian approach.
The coordinates on WS are distinguished: $\s$ -- compact coordinate,
$\s\in I$ for open string, $\s\in S^{1}$ for closed string;
$\tau\in{\bf R}^{1}$ -- non-compact coordinate, called evolution parameter.
The following denotations are introduced: 
$\dot x=\df x/\df\tau,\ x'=\df x/\df\s,\ p^{\mu}=p_{\tau}^{\mu}$,
so that the action of string theory (area) can be written as
$A=\int d\tau\int d\s\;\L,$ where 
$\L=\sqrt{(\dot xx')^{2}-\dot x^{2}x'^{2}}$ is called density of Lagrangian. 
Poisson brackets are introduced as
$\{x_{\mu}(\s,\tau),p_{\nu}(\tilde\s,\tau)\}=g_{\mu\nu}\delta(\s-\tilde\s)$,
where $g_{\mu\nu}=\mbox{diag}(+1,-1,...,-1)$ 
is metric tensor and $\delta()$ is Dirac's function.
The problem is stated to find a solution $x(\s,\tau),p(\s,\tau)$
starting from the initial data $x(\s,0),p(\s,0)$. 
This evolution is described by a system of autonomous 
differential equations, simultaneous by $\tau$
(so that $\tau$-dependence is usually omitted, implying that all 
variables are estimated at the same value of evolution parameter). 

String theory is Hamiltonian theory
with the 1st class constraints \cite{Dirac,Pronko}, this means the following.
The density of canonical Hamiltonian is defined by Legendre 
transformation as $\H_{c}=\dot xp-\L$. Substitution of $p$-definition 
in terms of $x',\dot x$ vanishes canonical Hamiltonian $\H_{c}=0$,
and additionally creates the following identities:
$\Phi_{1}=x'p=0,\ \Phi_{2}=x'^{2}+p^{2}=0$.
Appearance of these identities, called Dirac's constraints,
is related with the symmetry of the action under the group
of reparametrizations (right diffeomorphisms) of WS, i.e. smooth invertible
mappings $(\s,\tau)\to(\tilde\s,\tilde\tau)$. 
Usually consideration of theories 
with constraints starts in extended phase space $(x,p)$,
where Hamiltonian is defined as linear combination of constraints,
in our case $H=\int d\s(V_{1}\Phi_{1}+V_{2}\Phi_{2})$.
Here the coefficients $V_{1,2}(\s)$ are arbitrary 
and called Lagrangian multipliers. On the surface $\Phi_{i}=0$
the Hamiltonian vanishes\footnote{This scalar Hamiltonian
does not coincide with the energy, which in relativistic theories 
is given by a component of momentum $p_{0}$.}, 
however its derivatives do not vanish and create Hamiltonian vector field:
$$\dot x_{\mu}(\s)=\delta H/\delta p_{\mu}(\s),\ 
\dot p_{\mu}(\s)=-\delta H/\delta x_{\mu}(\s).$$
This field is tangent to the surface $\Phi_{i}=0$ due to the fact that
the Poisson brackets $\{\Phi_{i}(\s),\Phi_{j}(\tilde\s)\}$
vanish on the the surface $\Phi_{i}=0$, in this case the constraints
are said to be of the 1st class. Phase trajectory, integrated
from such field, belongs to the surface $\Phi_{i}=0$,
and its projection to coordinate space $\{x\}$ gives a solution
of Lagrange-Euler equations.
In string theory $\Phi_{i}$-terms of Hamiltonian generate
infinitesimal shifts of points in tangent directions to the WS:
$\Phi_{1}$ generates the shifts $\delta x\sim x'$, 
while $\Phi_{2}$ generates $\delta x\sim p\perp x'$, see fig.3.
Together the constraints generate all possible reparametrizations 
of WS (connected component of right diffeomorphisms group).

\begin{figure}\label{f3}
\begin{center}
~\epsfysize=1.5cm\epsffile{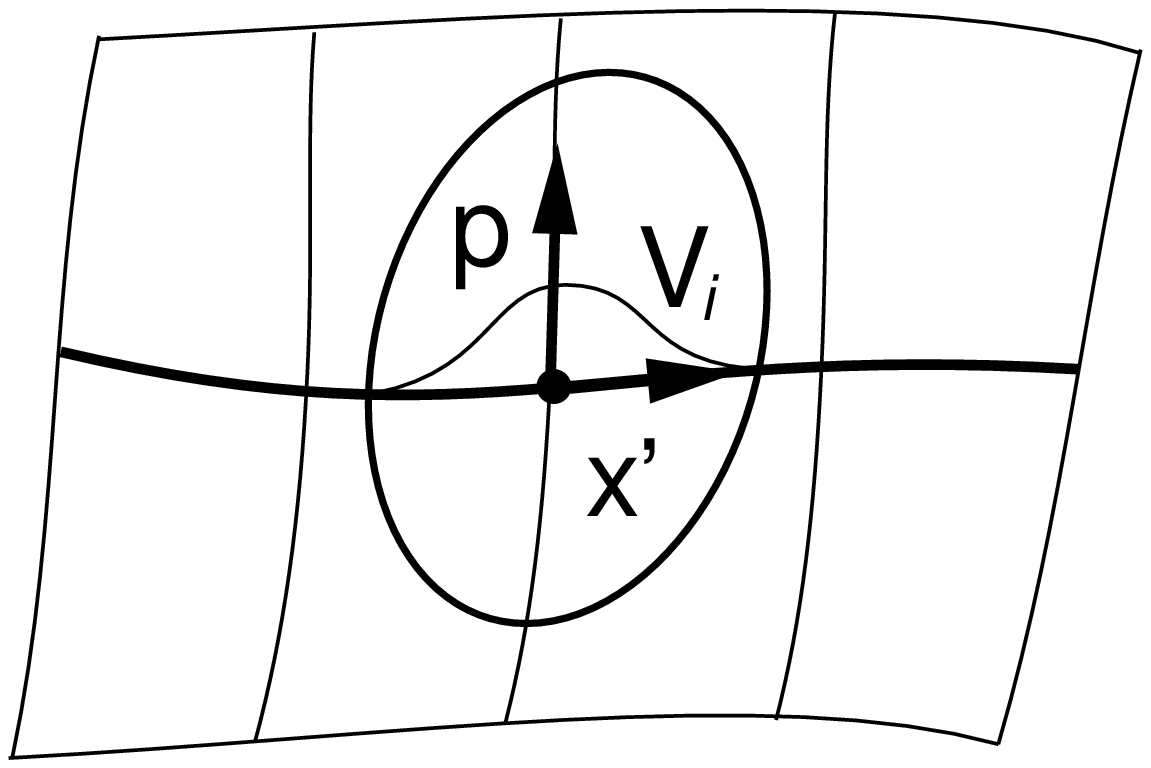}

\vspace{1mm}
\fignum Constraints generate reparametrizations.
\end{center}
\end{figure}

Coefficients $V_{1,2}$ influence only the parametrization of WS,
a choice $V_{1}=0,\ V_{2}=1$ corresponds to conformal parametrization
$\dot xx'=0,\ \dot x^{2}+x'^{2}=0$. This choice linearizes 
Hamiltonian equations. The equations were solved
by different methods in \cite{Pronko,Vladimirov,Brink,exotic},
the resulting solutions allow the following geometric representation.

\section{Structure of solutions: open and closed strings}

Reconstruction of WS is based on the concept of {\it supporting curves}.
Let's consider a curve $Q(\s)$ in Minkowski space
with the following properties:
(1) periodicity: $Q(\s+2\pi)=Q(\s)+2P$;
(2) light-likeness: $Q'^{2}=0$.

\begin{figure}\label{f4}
\begin{center}
~\epsfysize=2.5cm\epsffile{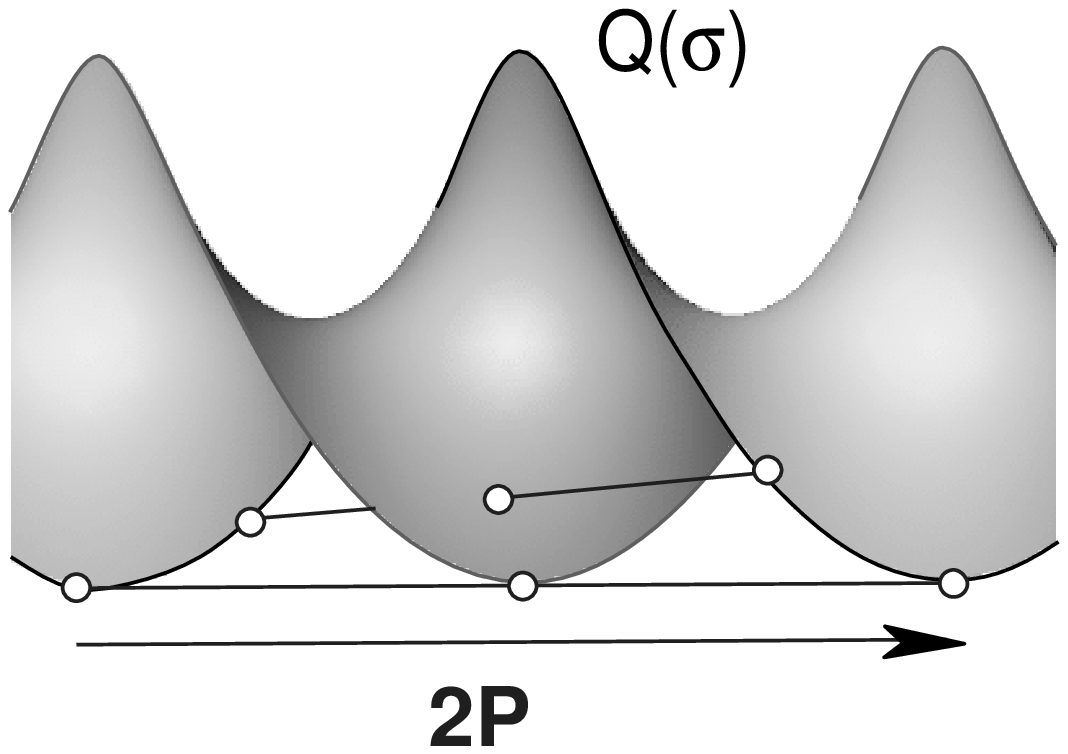}

\fignum WS of open string.
\end{center}
\end{figure}

\vspace{1mm}
WS of open string can be reconstructed by this curve as follows:
$x(\s_{1},\s_{2})=(Q(\s_{1})+Q(\s_{2}))/2,\ \s_{1}\leq\s_{2}\leq\s_{1}+2\pi$,
see fig.4. The obtained surface has two edges,
one coincident with $Q(\s)$, another one is $Q(\s)+P$.
Translation by vector $P$ transforms the WS to itself
with the edges interchanged.

\begin{figure}\label{f4b}
\begin{center}
~\epsfysize=2cm\epsffile{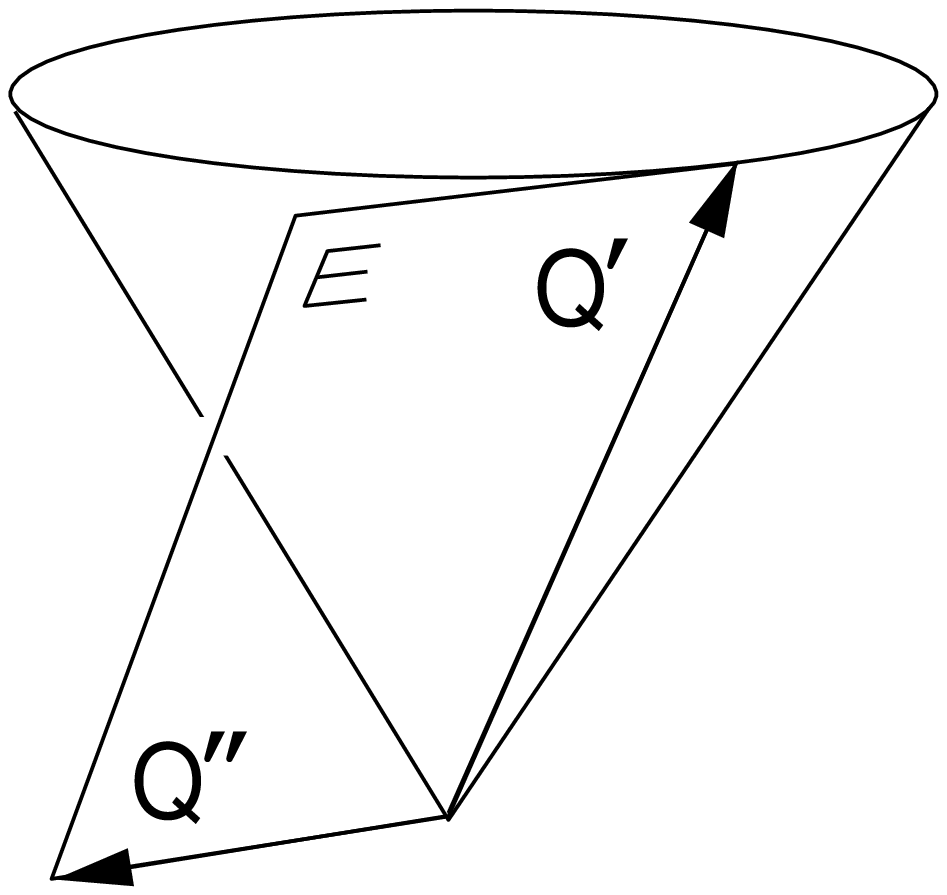}

\vspace{1mm}
\fignum Structure of WS near the edge.
\end{center}
\end{figure}

\noindent {\bf Theorem 2} (edge of WS): any vector from the
tangent plane to open WS on its edge is orthogonal to $Q'$.
This plane is tangent to the light cone in the direction of $Q'$
(such planes are called {\it isotropic}). As a consequence,
end points of open string move at light velocity
perpendicularly to the direction of the string in these points.

\vspace{2mm}
To construct the WS of closed string, one should consider two
supporting curves $Q_{1,2}(\s)$. Both should be light-like:
$Q_{1,2}'^{2}=0$ and periodical with the same period: 
$Q_{1,2}(\s+2\pi)=Q_{1,2}(\s)+P$. The WS is given by the formula:
$x(\s_{1},\s_{2})=(Q_{1}(\s_{1})+Q_{2}(\s_{2}))/2,\ 
\s_{1}\leq\s_{2}\leq\s_{1}+4\pi$.

Variables $\s_{i}$ introduce light-like coordinates on WS:
$(\df_{i}x)^{2}=0$, related with Hamiltonian coordinates
as $\s_{1,2}=\tau\pm\s$. The supporting curves can be reconstructed 
from initial data: coordinate $x$ and density of momentum $p$ 
on the string at value $\tau=0$ by the formulae
\begin{eqnarray}
a)&&Q(\s)=x(\s)+\int_{0}^{\s}d\tilde\s \;
p(\tilde\s),\ \mbox{for open string;}\label{Q}\\
b)&&Q_{1,2}(\s)=x(\pm\s)\pm\int_{0}^{\pm\s}d\tilde\s \;
p(\tilde\s),\ \mbox{for closed string.}\nn
\end{eqnarray}

In the last formula the curve $Q_{1}$ is given by
upper choice of all signs, while $Q_{2}$ is given by lower signs.
The functions $x(\s),p(\s)$ are continued to the 
whole axis of $\s$, see \cite{Brink}: 
as $2\pi$-periodical functions in (\ref{Q}b)
and as even $2\pi$-periodical functions in (\ref{Q}a). 

\vspace{1mm}\noindent{\it Remark}: 
Transition to complex variables $\tau\to i\tau,\ p\to ip$ 
in the above formulae gives a solution of Bj\"orling's problem 
\cite{Bjorling} for minimal surfaces in Euclidean space
(find a minimal surface, passing through a given curve and 
tangent to a given vector field on the curve).

\vspace{1mm}
The inverse formulae are:

\vspace{1mm}
\noindent$a)\ x(\s)=(Q(\s)+Q(-\s))/2,\
p(\s)=(Q'(\s)+Q'(-\s))/2,$

\noindent$b)\ \ x(\s)=(Q_{1}(\s)+Q_{2}(-\s))/2,\
p(\s)=(Q'_{1}(\s)+Q'_{2}(-\s))/2.$

\vspace{1mm}
From (\ref{Q}ab) we see that the vector $P$ 
coincides with the total energy-momentum
of the string. In particular case $Q_{1}=Q_{2}$ the WS of closed string
degenerates to 2-folded WS of open string. In this case
the periods of supporting curves are $P_{closed}=2P_{open}$,
as a result, each of two coincident sheets of resulting open string
receives a half of energy-momentum of original closed string.

Subspace, orthogonal to $P$, forms the center-of-mass frame (CMF).
In projection to CMF supporting curves become closed.
The curves, whose temporal component $Q^{0}(\s)$ {\it is not}
monotonous function, will be considered in the next section.
Here we consider supporting curves, satisfying at each point
the condition $Q'^{0}>0$. Such curves can be restored by their 
projection to CMF:
$$Q_{0}'=|\vec Q'|\ \Rightarrow\ Q_{0}(\s_{1})-Q_{0}(0)=
\int\nolimits_{0}^{\s_{1}}|\vec Q'(\s)|d\s=L(\s_{1}),$$
$L(\s)$ is
the arc length of the curve $\vec Q(\s)$ between the points $\vec Q(0)$ and
$\vec Q(\s)$. The total length of the curve $\vec Q(\s)$ is equal to
$2\sqrt{P^{2}}$ for open string and $\sqrt{P^{2}}$ for closed string.
One can parameterize the curve $\vec Q(\s)$ by its length:
$\s=2\pi L/L_{tot}$, then $Q_{0}'=|\vec Q'|=
L_{tot}/2\pi$. This parametrization is called Rohrlich gauge 
\cite{Rohrlich}. 

At the value of space-time dimension $d=3,4$ the strings have
stable singular points. Their appearance has the following reason.

\begin{figure}\label{f5}
\begin{center}
~\epsfxsize=8cm\epsffile{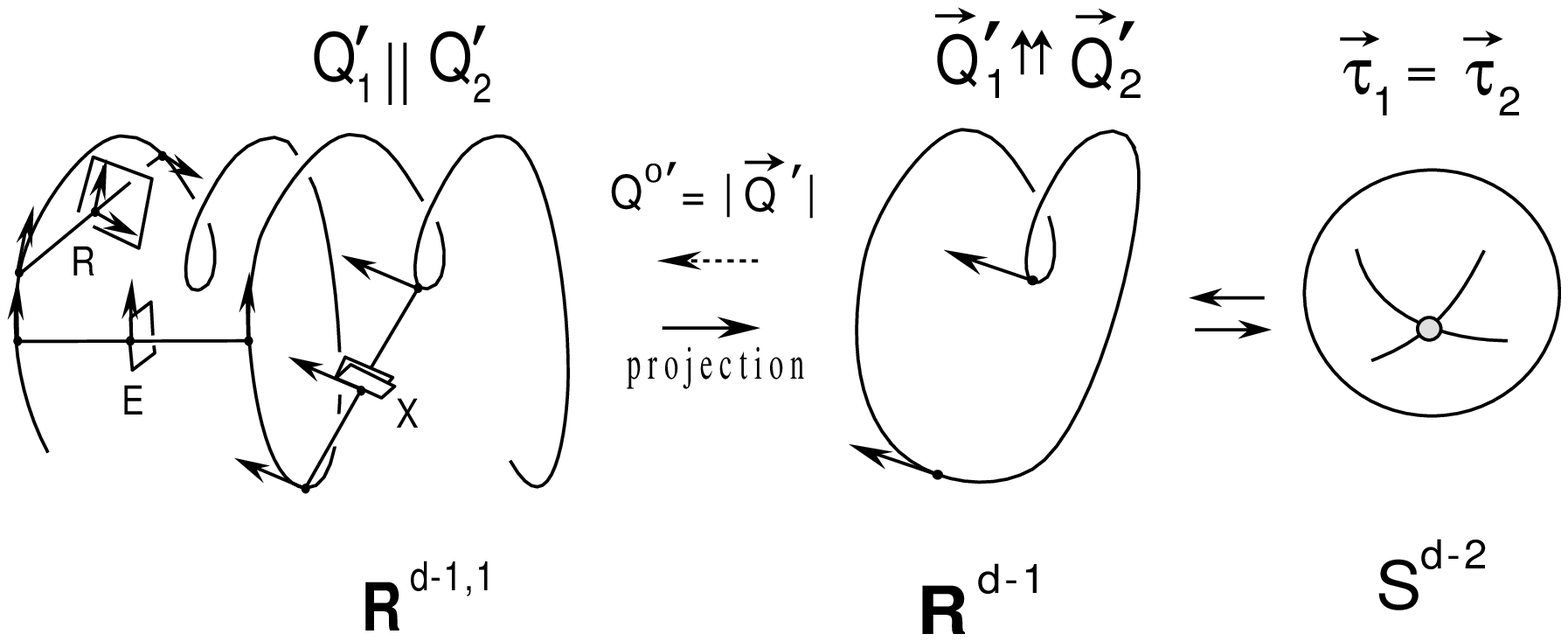}

\fignum Singularities of WS correspond to parallel
tangent vectors on supporting curve.
R -- regular point, E -- edge, X -- internal singular point.
\end{center}
\end{figure}

Tangent vectors to WS in a point $x(\s_{1},\s_{2})$
are $Q'(\s_{1})$ and $Q'(\s_{2})$, i.e. tangent vectors 
to the supporting curve in corresponding points. 
If these two vectors are linearly independent, 
they define tangent plane to WS,
and this point is regular (point R on \fref{f5}). 
Otherwise, if these vectors are parallel,
the point is singular. The point E with $\s_{2}-\s_{1}=0$ or $2\pi$
belongs to the edge of WS. The point X with $0<\s_{2}-\s_{1}<2\pi$
is internal singular point. 

\begin{figure}\label{f6}
\begin{center}
~\epsfysize=1.5cm\epsffile{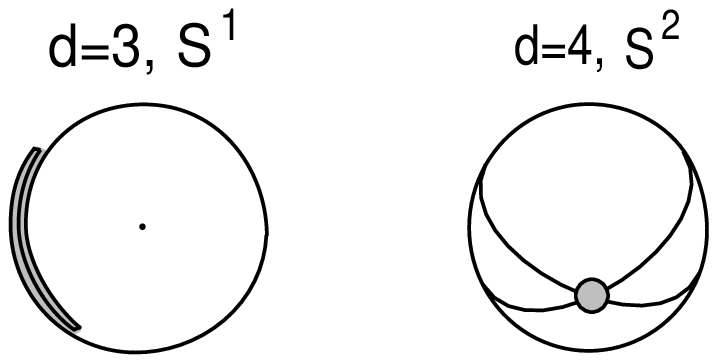}

\vspace{1mm}
\fignum Self-intersection of hodograph $\vec\tau(\s)$.
\end{center}
\end{figure}

For supporting curves with $Q'^{0}>0$ the linear dependence 
of vectors $Q'_{1}$ and $Q'_{2}$ in the space-time
is equivalent to the coincidence of unit tangent vectors
$\vec\tau=\vec Q'/|\vec Q'|$ to {\it the projection} of supporting curve
to CMF. Therefore, singularities of WS correspond to the points
of self-intersection of hodograph $\vec\tau(\s_{1})=\vec\tau(\s_{2})$
(for closed string -- intersection of two hodographs 
$\vec\tau_{1}(\s_{1})=\vec\tau_{1}(\s_{2})$). Hodographs belong to
a sphere $S^{d-2}$: to a circle for $d=3$ and to 2-dimensional sphere
for $d=4$. As a result, for $d=3$ the intersection of hodographs
is 1-dimensional set, and for $d=4$ the transversal intersections
are located in isolated points, see \fref{f6}. 
In the case $d>4$ the intersections can be removed by small variations
of hodographs on the sphere $S^{d-2}$.

From (\ref{Q}) we derive a formula for linear density 
of energy-momentum on the string:
$$dp_{0}/dl=2\gamma,\ d\vec p/dl=(\vec\tau_{1}+\vec\tau_{2})\gamma,\
\gamma=(2(1-\vec\tau_{1}\vec\tau_{2}))^{-1/2}.$$
In the singular points and on the edge of open WS $\gamma\to\infty$,
therefore in these points the linear density of energy-momentum
tends to infinity.

\begin{figure}\label{f7}
\begin{center}
~\epsfxsize=8cm\epsffile{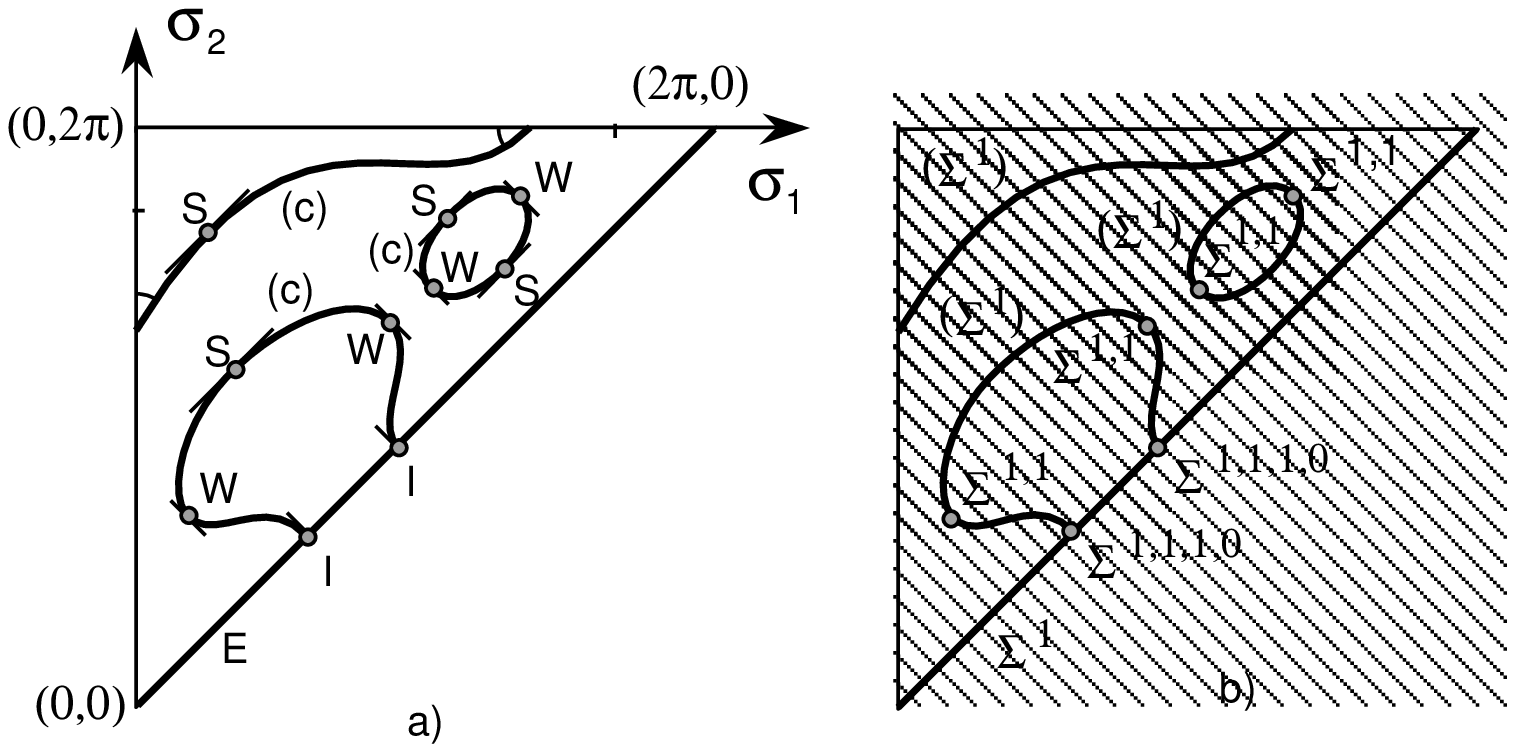}

\fignum a) singular lines on parametric plane (scheme); \\
b) classification of singularities according to degeneration
of \\ the 1st differential \cite{Arnold}.
\end{center}
\end{figure}

\paragraph*{Case ${\bf d=3}$:} representing 
$\vec\tau(\s)=(\cos\ph(\s),\sin\ph(\s))$,
singularities are defined by the equation $\ph(\s_{1})=\ph(\s_{2}) \mod2\pi$.

For open string this equation should be solved inside a triangle
on $(\s_{1},\s_{2})$-plane, shown on \fref{f7}a,
and then continued to the whole WS using the trivial symmetries
$\s_{1}\leftrightarrow\s_{2},\ \s_{1,2}\to\s_{1,2}+2\pi k,\ k\in{\bf Z}$.
For closed string the equation $\ph_{1}(\s_{1})=\ph_{2}(\s_{2}) \mod2\pi$
should be solved in a square $0<\s_{1,2}<2\pi$, and continued by
$\s_{1,2}\to\s_{1,2}+2\pi k$. Typically, for supporting curves 
in general position, solutions of these equations form smooth curves
on parametric plane, denoted on  \fref{f7}a as (c), for which the
following behavior is possible:

\vspace{1mm}
$\bullet$ curves extend on the WS not reaching its edges;

$\bullet$ curves can form closed loops;

$\bullet$ for open strings the curve (c) can terminate on the edge.

\begin{figure}\label{f8}
\begin{center}
~\epsfxsize=6cm\epsffile{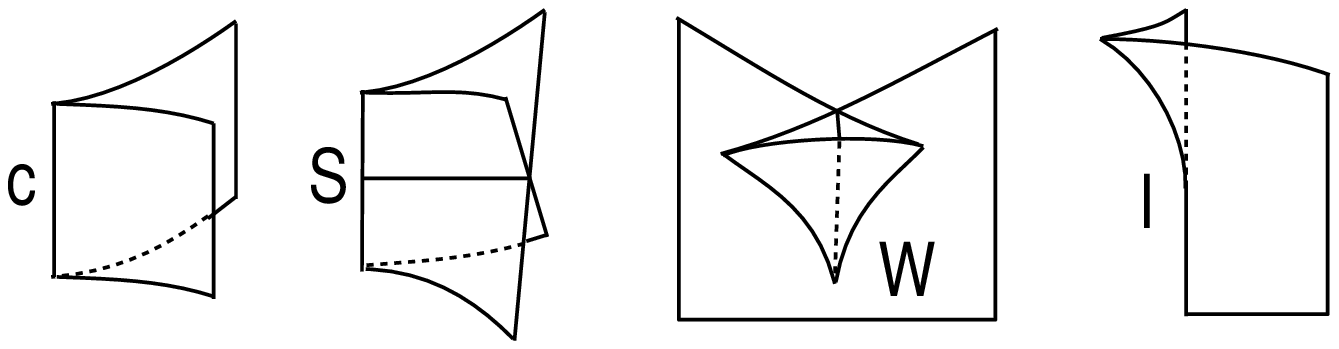}

\vspace{2mm}
\fignum Normal forms of singularities, $d=3$.
\end{center}
\end{figure}

\vspace{1mm}\noindent
In the last case the curve (c) enters to the edge at right angle,
due to the symmetry $\s_{1}\leftrightarrow\s_{2}$. Additionally, 
the points are marked on \fref{f7}a, where the curve (c)
is tangent to directions $d\s_{2}=\pm d\s_{1}$. The evolution
is performed by equal-time slices $x_{0}=Const$, which in Rohrlich gauge
correspond to $\s_{1}+\s_{2}=Const$. 
A structure of WS near singular points is shown on \fref{f8}:

\vspace{-2mm}
\begin{eqnarray}
E:&\quad& (v,u^{2},0)\nn\\
c:&\quad& (v,u^{2},u^{3})\nn\\
S:&\quad& (v,u^{2},vu^{3})\nn\\
W:&\quad& (v^{2}+2u,v^{3}+3vu,v^{4}+4v^{2}u)\nn\\
I:&\quad& (v^{2}+u,v^{4}+2v^{2}u,v^{6}+3v^{4}u)\nn
\end{eqnarray}

\vspace{-2mm}

The normal forms are given in light coordinates $(t,y,x-t)$,
so that the direction $(1,0,0)$ is light-like, 
and together with the direction $(0,1,0)$ it defines the isotropic plane.

\vspace{1mm}\noindent
E: edge of WS;

\vspace{1mm}\noindent
c: equal-time slices of WS represent a cusp
$(0,u^{2},u^{3})$, moving in the space-time in light-like 
direction $(1,0,0)$, i.e. cusp moves at light velocity 
perpendicularly to its own direction $(0,1,0)$. 

\vspace{1mm}\noindent
S: line of self-intersection of the surface terminate on the cusp line,
in this point ``the wings'' of cusp pass through each other.

\vspace{1mm}\noindent
W: swallowtail \cite{Arnold}, a point of creation or annihilation 
of two cusps.

\vspace{1mm}\noindent
I: for open string the cusp can appear/disappear alone at the edge of WS,
in the point where this curve has inflection: $\ph'(\s_{I})=0$.
In other words, a cusp absorbed by the end
of the string inflects its trajectory.

\paragraph*{Case ${\bf d=4}$} ~~

\begin{figure}\label{f9}
\begin{center}
\parbox{1cm}{~\epsfxsize=1cm\epsffile{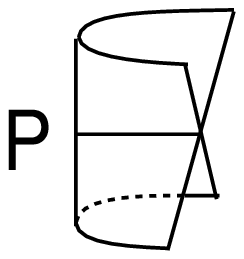}}\quad\quad~\parbox{4cm}{
\begin{eqnarray}
E:&\quad& (v,u^{2},0,0)\nn\\
P:&\quad& (v,u^{2},vu,u^{3})\nn
\end{eqnarray}
}

\vspace{2mm}
\fignum Normal forms of singularities, $d=4$.
\end{center}
\end{figure}

\noindent
Here we use light coordinates $(t,y,z,x-t)$.
The direction $(1,0,0,0)$ is light-like, 
and together with $(0,1,0,0)$, $(0,0,1,0)$ 
it defines an isotropic 3-dimensional hyperplane, tangent 
to light cone in 4 dimensions.

Equal-time slices of WS near point (P)
give a curve, which is smooth at $t\neq0$ 
and has a cusp $(0,u^{2},0,u^{3})$ at the instant of time $t=0$.
Projections to almost any\footnote{Exceptional directions of projection
belong to a plane $t=x$, they create higher order singularities.} 
3-dimensional subspace transform this singularity 
to {\it pinch point}, where the surface has a form 
of {\it Whitney umbrella}: $(v,u^{2},vu).$ 

The following theorems describe the transformation of WS to normal forms.
Let $\vec Q:\ S^{1}\to{\bf R}^{n}$ be $C^{\infty}$-smooth analytical
function with $\vec Q'(\s)\neq0$, i.e. smooth closed curve in ${\bf R}^{n}$.
Let $\s$ be a natural parameter (length) on the curve.

\vspace{1mm}\noindent
{\bf Definition} $(n=2)$: Let $\ph(\s)$ be polar angle of the vector 
$\vec Q'(\s)$. Let's denote $\ph_{i}=\ph(\s_{i})$.
We say that the curve is {\it in general position},
if in every pair of points where the equation $\ph_{1}=\ph_{2} \mod2\pi$ 
is satisfied, one of the following properties is valid:

(c) $\ph_{1}'\neq\pm\ph_{2}'$;

(S) $\ph_{1}'=\ph_{2}'\neq0,\ \ph_{1}''\neq\ph_{2}''$;

(W) $\ph_{1}'=-\ph_{2}'\neq0,\ \ph_{1}''\neq\ph_{2}''$;

(E) $\s_{1}=\s_{2},\ \ph_{1}'\neq0$;

(I) $\s_{1}=\s_{2},\ \ph_{1}'=0,\ \ph_{1}''\neq0,\ \ph_{1}'''\neq0$.

\noindent For two smooth closed curves $\vec Q_{1,2}$ in ${\bf R}^{2}$ 
let's denote $\ph_{i}=\ph_{i}(\s_{i})$.
We say that two curves are
{\it in general position to each other}
if in every pair of points where the equation 
$\ph_{1}=\ph_{2} \mod2\pi$ is satisfied, 
one of the properties (c),(S),(W) is valid.

\vspace{1mm}\noindent
{\bf Definition} $(n=3)$: For the curve in ${\bf R}^{3}$
let's denote $\vec Q_{i}'=\vec Q'(\s_{i})$.
The curve is in general position,
if in every pair of points 
where the equation $\vec Q_{1}'=\vec Q_{2}'$ is satisfied, 
one of the following properties is valid:

(E) $\s_{1}=\s_{2},\ \vec Q_{1}''\neq0$;

(P) $\vec Q_{1}'' \nparallel \vec Q_{2}'',\ 
|\vec Q_{1}''|\neq|\vec Q_{2}''|$.

\noindent For two curves in ${\bf R}^{3}$ let's denote 
$\vec Q_{i}'=\vec Q_{i}'(\s_{i})$. 
Two curves are in general position to each other if 
the property (P) is valid in every pair of points 
where the equation $\vec Q_{1}'=\vec Q_{2}'$ is satisfied.

\vspace{1mm}\noindent
{\bf Theorem 3} (general position):
curves in general position form an open everywhere dense set
in $C^{\infty}$-topology. 

\vspace{1mm}\noindent
{\bf Theorem 4} (normal forms): 
for supporting curves in general position
the singularities of the WS in their neighborhood 
can be transformed to one of the above written normal forms 
by LR-diffeomorphisms,
i.e. by smooth invertible transformations of parameters plane
$(\s_{1},\s_{2})\to(\tilde\s_{1},\tilde\s_{2})$ and 
Minkowski space $x\to\tilde x$.
The linear part of the transformation of Minkowski space,
defined by Jacoby matrix $\df\tilde x/\df x$,
maps light-like direction $Q_{1}'\|Q_{2}'$
to light-like direction $(1,0,..,0)$ and also maps
to each other the isotropic planes, related with these directions.

\vspace{1mm}
Theorem 3 means that curves in general position 
represent a general case in the set of all smooth curves.
Refinement in the theorem 4 relates the isotropic planes
in the image and pre-image spaces and
allows to write normal forms in light coordinates.

\paragraph*{${\bf\Sigma}$-classification \cite{Arnold}.}
$\Sigma$-class (Thom's symbol) of singularity 
is defined by a kernel of the 1st differential 
of considered mapping, in our case: solution of an equation
$Q'_{1}d\s_{1}+Q'_{2}d\s_{2}=0$, which in Rohrlich gauge
follows $d\s_{1}+d\s_{2}=0$. This kernel is 1-dimensional,
so that the curves (c), edges (E) and points (P)
are singularities of class $\Sigma^{1}$.
The lines $d\s_{1}+d\s_{2}=0$ are shown by hatching on \fref{f7}b.
The points (W), where the hatching is tangent to the curves (c),
correspond to singularities of class $\Sigma^{1,1}$
(cusp on the line of cusps). For points (I) Thom's symbol
is not defined, however in Boardman's refining 
scheme these points are classified as $\Sigma^{1,1,1,0}$.
Points (S) are of class $\Sigma^{1}$, i.e. are not distinguished
from surrounding points of cusp line (c) by $\Sigma$-classification.

\begin{figure}\label{f10}
\begin{center}
~\epsfxsize=8cm\epsffile{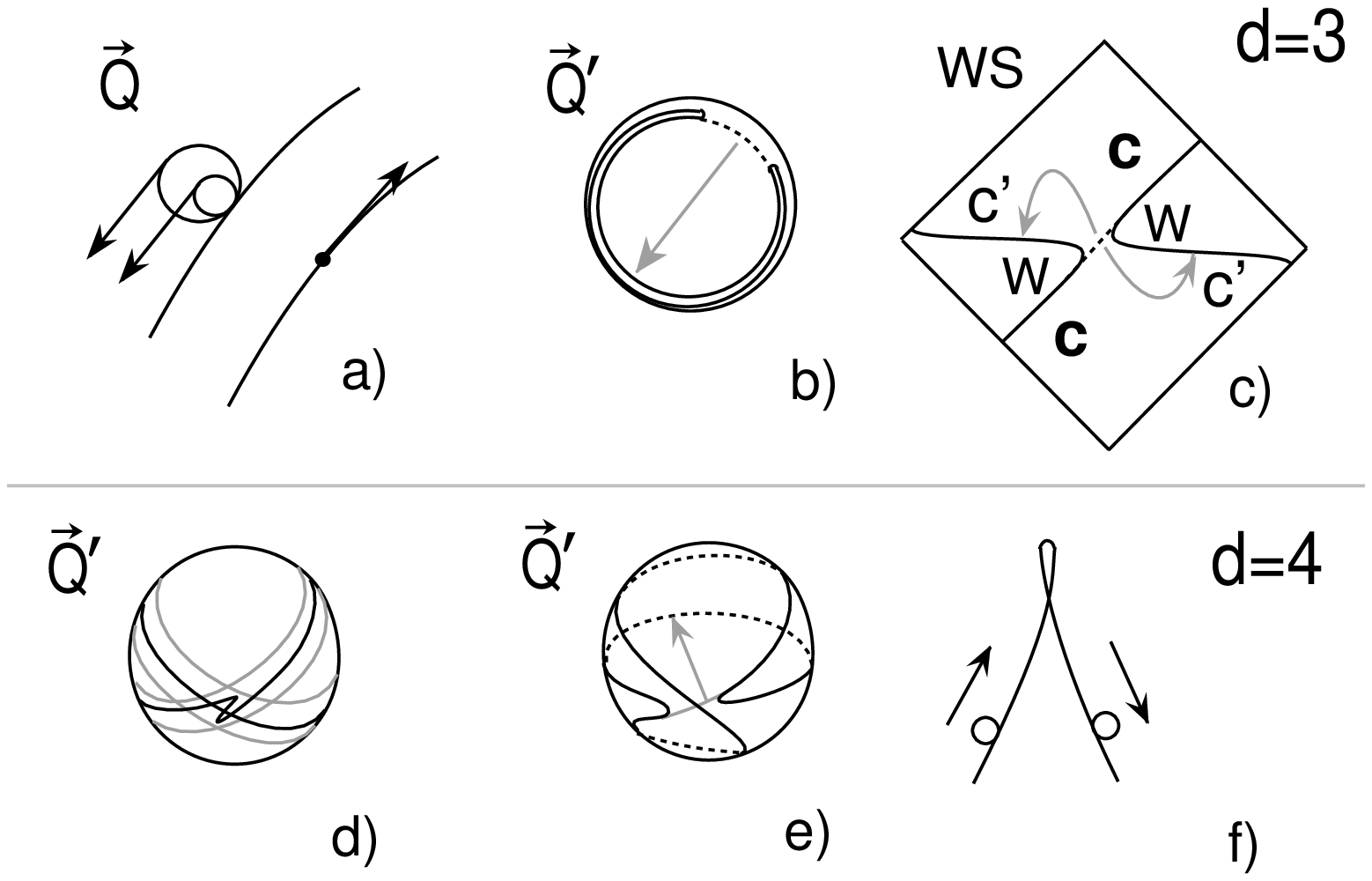}

\fignum Deformations, changing structure of singularity.
\end{center}
\end{figure}

\paragraph*{Stability.} Let's call the singularity of WS {\it weak stable} 
(W-stable), if all WS in $\epsilon$-vicinity of the given WS$_{0}$ 
have a singularity. The singularity is called LR-stable \cite{Arnold}, 
if all WS in $\epsilon$-vicinity of WS$_{0}$ 
can be transformed to WS$_{0}$ by LR-diffeomorphism
(in this case the singularity will be certainly W-stable).
The described singularities are LR-stable if $\epsilon$-vicinity of WS
is defined in $C^{\infty}$-topology: 
$|\delta\vec Q^{(n)}|<\epsilon_{n},\ n=0,1,2...$
The singularities are W-stable in $C^{1}$-topology: 
$|\delta\vec Q'|<\epsilon$ (in this case they already are not
LR-stable, example is shown on \fref{f10}d: 
near the point of transversal self-intersection of hodograph
$\vec Q'(\s)$ in its $|\delta\vec Q'|<\epsilon$ -vicinity
$C^{1}$-small but $C^{2}$-large variations exist, 
changing the structure of self-intersection).
$C^{1}$-small variations of supporting curves are
equivalent to variations of initial data, small in the sense
$|\delta x'|<\epsilon,\ |\delta p|<\epsilon$. From physical point of view
it is also interesting to consider $C^{0}$-small variations:
$|\delta\vec Q|<\epsilon$, equivalent to small variation
of coordinate $|\delta x|<\epsilon$ and small change of integrals 
$|\int d\s\delta p|<\epsilon$, taken over finite segments of the
string. The variations of supporting curve, 
which are $C^{0}$-small but $C^{1}$-large,
correspond to vanishing loops, displayed in \fref{f10}a.
These loops can eliminate the coincidence of tangent vectors,
throwing the hodograph to an opposite side of a circle, \fref{f10}b,
therefore removing the singularity. More detailed consideration
shows that in the case $d=3$ this process generates new
cusps $(c')$ and swallowtails $(W)$, see \fref{f10}c, so that the singularity
can be removed in small region of deformation,
but not completely from the whole WS. The pinch points
can be completely eliminated by $C^{0}$-small deformations, see \fref{f10}e.
This process is followed by a creation of a small loop on the string,
\fref{f10}f, 
which propagates at light velocity and passes through the point (P),
preventing the appearance of the instantaneous cusp there. 
In this region of the string the linear density of energy-momentum 
is large but is not infinite.

It's also interesting to study the properties of singularities in
Fourier representation \cite{Brink}, where the coefficients of 
expansion $Q'(\s)=\sum a_{n}e^{in\s}$ are used
as coordinates in the phase space of the string.
In finite dimensional subspaces $\{a_{n},n\leq N;\ a_{n}=0,n>N\}$
subsets with $\vec Q'\neq0$ and only transversal intersections of hodograph
$\vec\tau=\vec Q'/|\vec Q'|$ are open, therefore
the singularities are W-stable;
and because in such sets $\vec\tau$ are analytical,
the singularities are LR-stable. For creation of the vanishing loops 
the infinite number of Fourier coefficients is necessary. 
In the space of all $\{a_{n}\}$ the definitions of $C^{0}$- and $C^{1}$-small
variations can be reproduced by certain requirements 
for the rate of descent of Fourier coefficients.

This table presents several settings of topology
and the resulting stability type of singularities on WS:

\begin{center}
\begin{tabular}{|c|c|c|c|}\hline
Topology&$d=3$&$d=4$&$d>4$\\\hline
$|\delta\vec Q^{(n)}|<\epsilon_{n}\ (C^{\infty})$&LR&LR&---\\
$|\delta\vec Q'|<\epsilon\ (C^{1})$&W&W&---\\
$|\delta\vec Q|<\epsilon\ (C^{0})$&W&---&---\\\hline
$|\delta a_{n}|<\epsilon;\ a_{n}=0,n>N$&LR&LR&---\\
$|\delta a_{n}|<\epsilon/n^{p},p>1\ (C^{1})$&W&W&---\\
$|\delta a_{n}|<\epsilon/n^{p},p>0\ (C^{0})$&W&---&---\\\hline
\end{tabular}
\end{center}

WS can also have stable self-intersections, 
whose properties are the same as for generic surfaces.
The following tables summarize all stable singularities 
for WS and generic surfaces (in $C^{\infty}$-topology).

\vspace{-1mm}
\begin{center}
Self-intersections

\vspace{1mm}
\begin{tabular}{|c|c|c|c|}\hline
Surface &$d=3$&$d=4$&$d>4$\\\hline
generic, WS&lines&points& ---\\\hline
\end{tabular}

\vspace{2mm}
Others

\vspace{1mm}
\begin{tabular}{|c|c|c|c|}\hline
Surface &$d=3$&$d=4$&$d>4$\\\hline
generic&pinch points&---& ---\\\hline
&cusp lines and && \\
WS&their singulari-&pinch points&---\\
&ties (S),(W),(I)&&\\\hline
\end{tabular}
\end{center}

\vspace{-2mm}
\paragraph*{Global structure of singularities:} {\it ($d=3$, closed string)}

\noindent
Further consideration will be done in CMF.

\vspace{1mm}\noindent
{\bf Theorem 5} (presence of singularities) \cite{ECHAQ}:
all WS of closed string in 3-dimensional Minkowski space necessarily 
have singularities.

\vspace{1mm}\noindent
{\it Remark}: at certain conditions (central symmetry 
of supporting curves) the WS has singularity of type ``collapse'',
where the string for an instance of time shrinks to a point.
Example: $Q_{1,2}(\s)=(\s,\cos\s,\pm\sin\s)$, the supporting
curves, whose projection to CMF are two oppositely oriented circles.
This singularity is unstable: small variations of the curves 
unfold it to a small closed cusp line.

\vspace{1mm}\noindent
The velocity of cusp $\vec v$ is orthogonal to its direction $\vec k$.

\vspace{-1mm}\noindent
\parbox{4.1cm}{
{\bf Definition:} 

{\it topological charge of cusp} is a number $c$, equal to $+1$, if
a rotation from $\vec v$ to $\vec k$ is {\it counterclockwise}; 
and equal to $-1$ if this rotation is clockwise.
}\parbox{4.2cm}{
\begin{figure}\label{f11}
\begin{center}
~\epsfysize=2cm\epsfxsize=4.6cm\epsffile{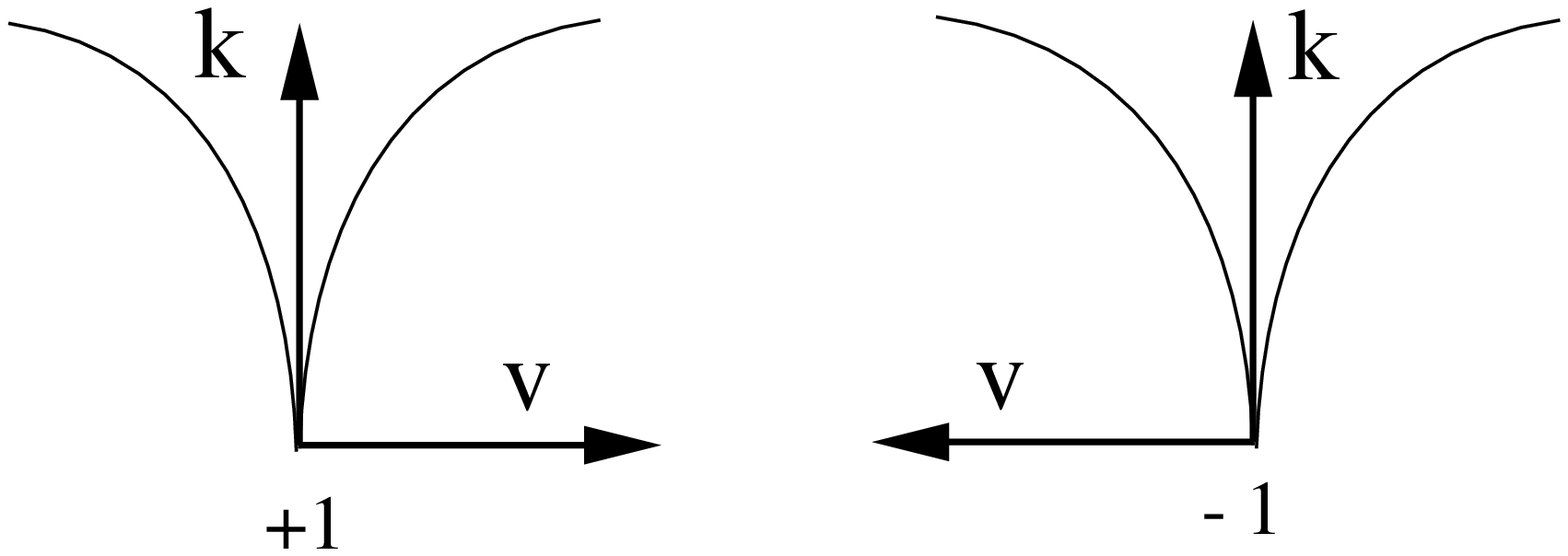}

\fignum Topological charge.
\end{center}
\end{figure}
}

\vspace{-1mm}\noindent
{\bf Theorem 6} (conservation of topological charge):

\noindent
Total topological charge of the string, equal to the sum 
of topological charges of all cusps, is constant in time and 
equals $n_{1}+n_{2}$. Here $n_{i}$ are the numbers of 
revolutions of vectors $\vec Q'_{i}(\s)$ in complete passage
around supporting curves ($n_{i}>0$, if revolutions are counterclockwise;
$n_{i}<0$, if they are clockwise). 

\vspace{2mm}\noindent
{\bf Theorem 7} (permanent regime):
Let supporting curves 

\noindent\parbox{4.1cm}{
$\vec Q_{i}(\s)$ have no inflection points. Let $\sgn n_{1}=\sgn n_{2}$. 
In this case cusps do not collide and topological charges 
of all cusps have the same sign equal to $\sgn n_{1,2}$.
As a result, {\it the number of cusps} on the string is constant in time
and equals $|n_{1}+n_{2}|$.
}\quad\parbox{4cm}{
\begin{figure}\label{f12}
\begin{center}
~\epsfysize=2cm\epsfxsize=4cm\epsffile{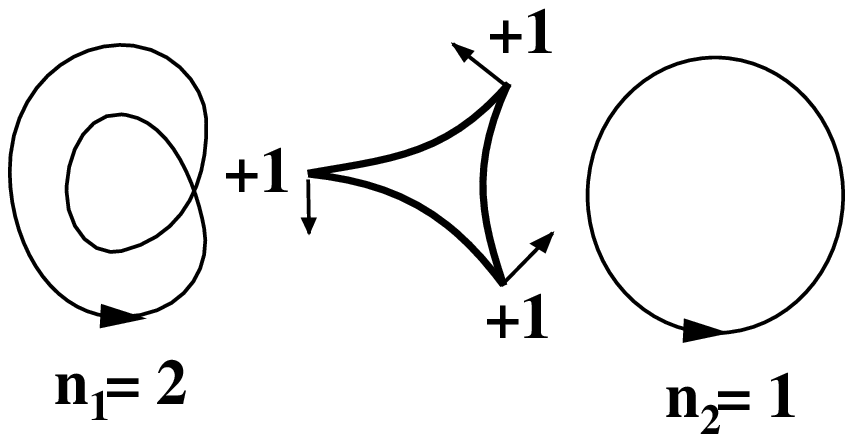}

\fignum Permanent regime.
\end{center}
\end{figure}
}

\vspace{2mm}\noindent
{\it Remark:} curves with $n_{i}=0$ (e.g. figure ``{\large $\infty$}'')
necessarily have inflection points and violate
permanent conditions. Thus, these conditions imply $|n_{i}|\geq1$, and
the strings under permanent conditions always have $N\geq2$ cusps.

\begin{figure}\label{f14}
\begin{center}
~\epsfxsize=4cm\epsffile{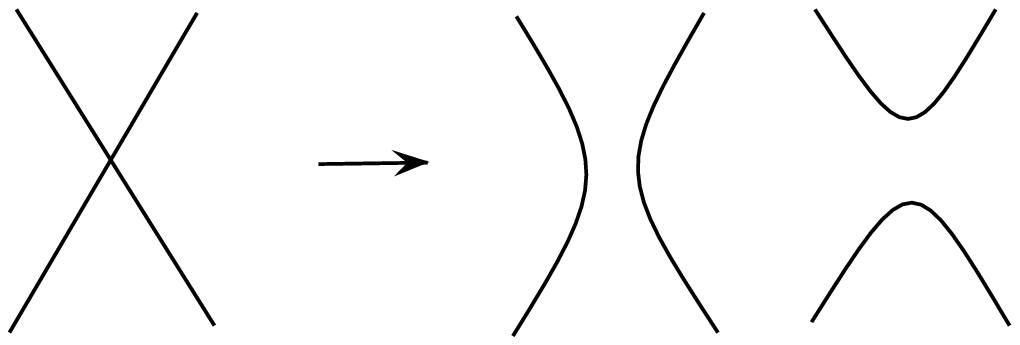}

\vspace{1mm}
\fignum Intersection of cusp lines is unstable.
\end{center}
\end{figure}

\vspace{1mm}\noindent
\noindent{\bf Theorem 8} (collision of cusps):
intersection of cusp lines is unstable, i.e. small variations
of supporting curves transform it either to scattering or to 
annihilation/creation mode (\fref{f14}). Cusps are created/annihilated
by pairs in singular points of type W. At the moment of creation 
cusps have equal velocities and opposite directions, 
so that conservation of topological charge
is valid: $(+1,-1)\leftrightarrow0$.

\begin{figure}\label{f16}
\begin{center}
~\epsfxsize=8cm\epsffile{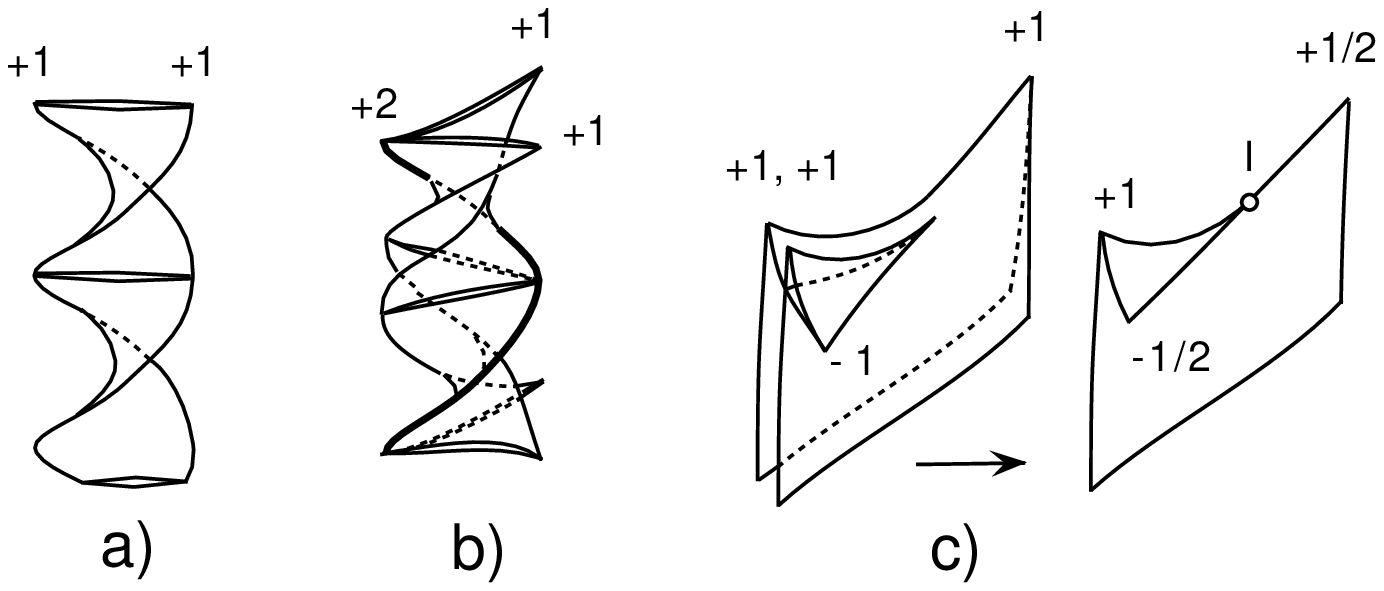}

\fignum Folding of WS: closed $\to$ open.
\end{center}
\end{figure}

\vspace{2mm}\noindent
{\it Open strings} can be free of singularities. Example 
\cite{straight}: the supporting curve $\vec Q(\s)$ is a circle, 
WS is helicoid \fref{f4}, the string in CMF is a straight line,
rotating at constant angular velocity. Open string is a degenerate case
of closed one, so that analogous theorems for open strings 
can be obtained in the limit $\vec Q_{1}\to\vec Q_{2}$. This limit maps
two cusps of closed string to the edges of WS, while other cusps
become 2-folded, see \fref{f16}a,b. 
From the obtained two coincident sheets only one represents the open
string, so that a half of the topological charges of singularities
should be taken, leading to the charge $\pm1$ for cusps and
$\pm1/2$ for the edges. Absorption of the cusp at the edge 
corresponds to the process \fref{f16}c:
$(+1,-1/2)\leftrightarrow+1/2$.

\vspace{2mm}\noindent{\bf Theorem 6'}:
total topological charge of open string, equal to the sum 
of topological charges of cusps and edges, is constant in time and 
equals $n$, the number of 
revolutions of vector $\vec Q'(\s)$ in the passage
around supporting curve.

\vspace{2mm}\noindent{\bf Theorem 7'}: if the supporting curve
$\vec Q(\s)$ has no inflection points,
the cusps do not collide and do not reach the edges,
and the topological charges 
of all cusps and edges have the same sign, equal to $\sgn n$.
The total number of cusps in this case is constant in time
and equals $|n|-1$.

\vspace{2mm}\noindent{\bf Theorem 8'}: 
there exists stable intersection of cusp line
with the edge of WS in the points of type I. This singularity
can be obtained in folding of the swallowtail with
the cusp line in the limit $\vec Q_{1}\to\vec Q_{2}$, 
as shown on \fref{f16}c. 

\vspace{3mm}
\noindent\parbox{5.9cm}{
{\it Remark:} the described local elements can be assembled
to more complicated patterns, particularly, the following
process can be found both on closed and open WS, see \fref{Z-proc}: 
a cusp (0) is initially present on the string, 
then the pair of cusps (1,2) appears,
one cusp from the pair annihilates the cusp (0),
while another one becomes the cusp (0) in the next 
period of evolution.
}~~~\parbox{2cm}{
\begin{figure}\label{Z-proc}
\begin{center}
~\epsfysize=2cm\epsffile{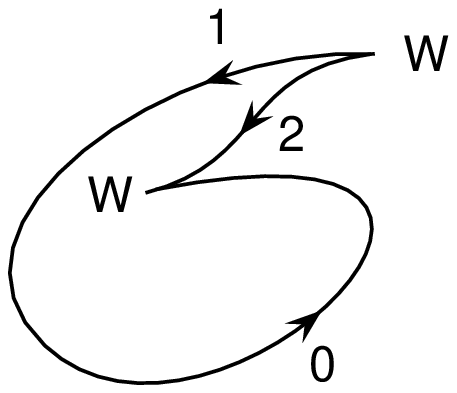}

\vspace{1mm}
\fignum Z-process.
\end{center}
\end{figure}
}

\vspace{2mm}
Several examples of singularities on strings at $d=3$ are shown
on \fref{col-plate}a-d.

\paragraph*{Global structure of singularities:} ($d=4$)

\noindent
Stable singularities are pinch points, periodically located on WS. 
In evolution instantaneous cusps appear on the string
in its passage through the pinch points, periodically at the same 
point in CMF. 3-dimensional projections of the WS of open string 
from 4-dimensional Minkowski space-time are shown on \fref{col-plate}e,f. 
Two types of singular points can be found in this figure:
$P,P',...$ -- singular points, existing on WS itself, 
which are projected to pinch points in projections 
to 3-dimensional space, such as $(xyz),\ (xyt)$,
shown in the figure; $Q$ -- pinch point, which appears only 
on a specific projection and therefore is not physically important.

For the pinch points the topological charge also can be introduced,
characterizing the behavior of singularities in continuous deformations
(homotopies) of the WS. For closed strings the transversal intersections
of two closed oriented curves $\vec Q'_{1,2}(\s)$ on the sphere $S^{2}$
are characterized by indices \cite{DNF}, equal to $+1/-1$ 
if a pair of tangent vectors $(\vec Q''_{1},\vec Q''_{2})$ 
has coincident/opposite orientation with a frame defining 
global orientation on the sphere $S^{2}$. Due to the theorems \cite{DNF},
the sum of all indices is invariant under homotopies
and for two closed curves on a sphere is always equal to zero.
Therefore, the number of pinch points counted in one period
of WS of closed string is even: $n=0,2,4,6...$ and in continuous
deformations of WS the pinch points appear/disappear pairwise:
$(+1,-1)\leftrightarrow0$.

\begin{figure}\label{f17}
\begin{center}
~\epsfxsize=7cm\epsffile{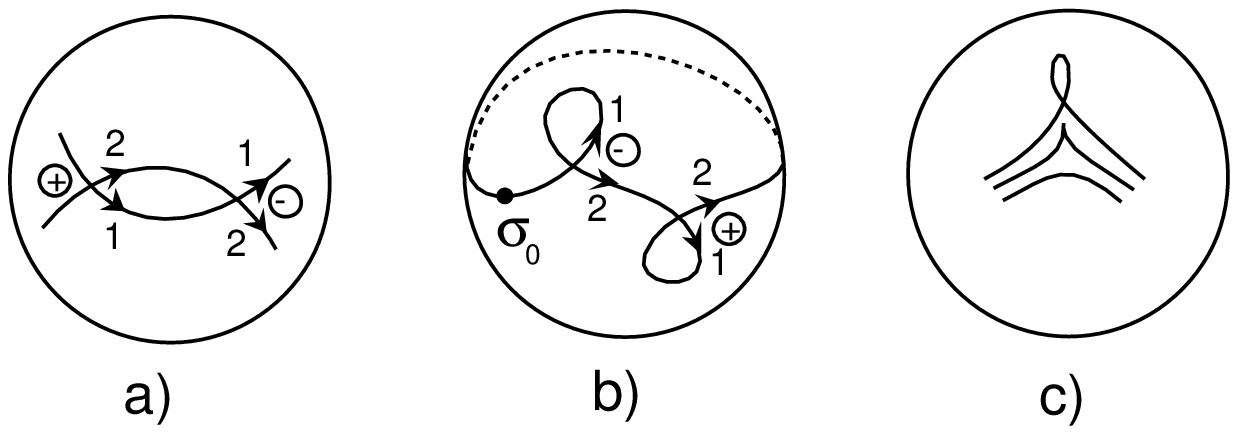}

\fignum Topological charge of pinch point.
\end{center}
\end{figure}

For open strings the points of self-intersection of the curve 
$\vec Q'(\s)$ can appear/disappear alone, followed by the creation
of cusp $\vec Q''(\s)=0$, see \fref{f17}c. 
This situation corresponds to the creation
of inflection point on the supporting curve $\vec Q(\s)$
in 3-dimensional space. On WS the pinch point moves to the edge
($\vec Q'_{1}=\vec Q'_{2},\ \s_{1}\to\s_{2}$) and disappears
in the inflection point of the edge. The pinch points can also
appear/disappear pairwise in internal regions of WS
during its continuous deformation, preserving a local characteristic,
introduced as follows \cite{DNF}. Let's fix on the curve $\vec Q'(\s)$
a point $\s_{0}$, not coincident with self-intersection.
Let's pass a curve starting from $\s_{0}$ and mark the tangent vectors
in intersection points: in the 1st passage through the intersection point
write 1 on the corresponding tangent vector, and in the 2nd passage
write 2, see \fref{f17}b. Then assign to the intersection point a number
$\pm1$ dependently on the orientation of the frame $(1,2)$. 
The sum of these numbers is called index of self-intersection 
(Whitney number) of the closed curve. This number
depends on a choice of $\s_{0}$ (in a passage of $\s_{0}$ 
through the point of self-intersection its index changes the sign),
however {\it the parity} of Whitney number does not
depend on $\s_{0}$ and is invariant under those homotopies
which do not create the cusps $\vec Q''(\s)=0$.

\section{Exotic solutions}

\begin{figure}\label{f18}
\begin{center}
~\epsfxsize=4cm\epsffile{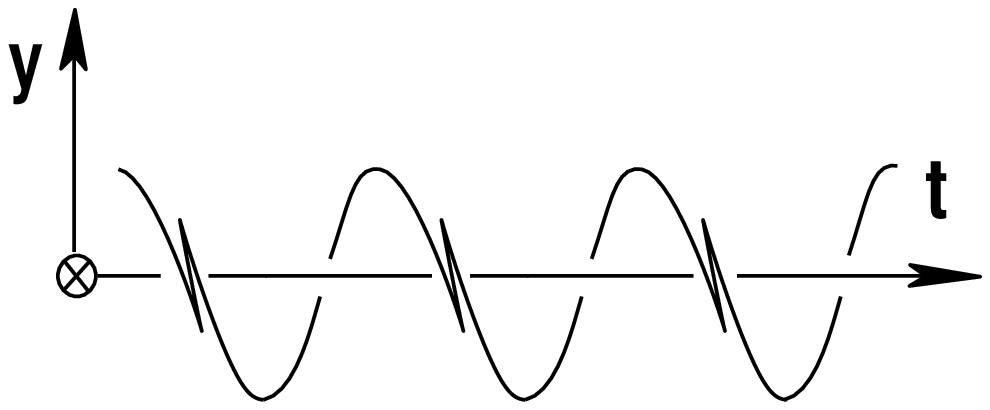}

\fignum Supporting curve, non-monotonous \\ in temporal direction.
\end{center}
\end{figure}

Solutions of this form correspond 
to the supporting curves,
whose temporal component $Q_{0}(\s)$ is non-monotonous function,
see \fref{f18}. Such curves can be explicitly constructed, specifying
tangent vector in the form $Q'(\s)=a_{0}(\s)(1,\vec n(\s)),$
where $|\vec n(\s)|=1$, $\vec n(\s)$ is $2\pi$-periodic function,
and $a_{0}(\s)$ is $2\pi$-periodic function of variable sign.
Corresponding WS is shown on \fref{col-plate}g.
On this figure (cABh) is a supporting curve, which has two cusps A,B. 
These cusps induce cusp lines on WS: (fRAd) and (gBRe), 
which separate the WS into  a number of pieces. 
Here R=(A+B)/2.

Equal-time slices of this WS contain disconnected parts. 
There is a long string, which is permanently present in the system.
Additionally, the following processes occur:

\begin{figure}\label{f19}
\begin{center}
~\epsfysize=5cm\epsffile{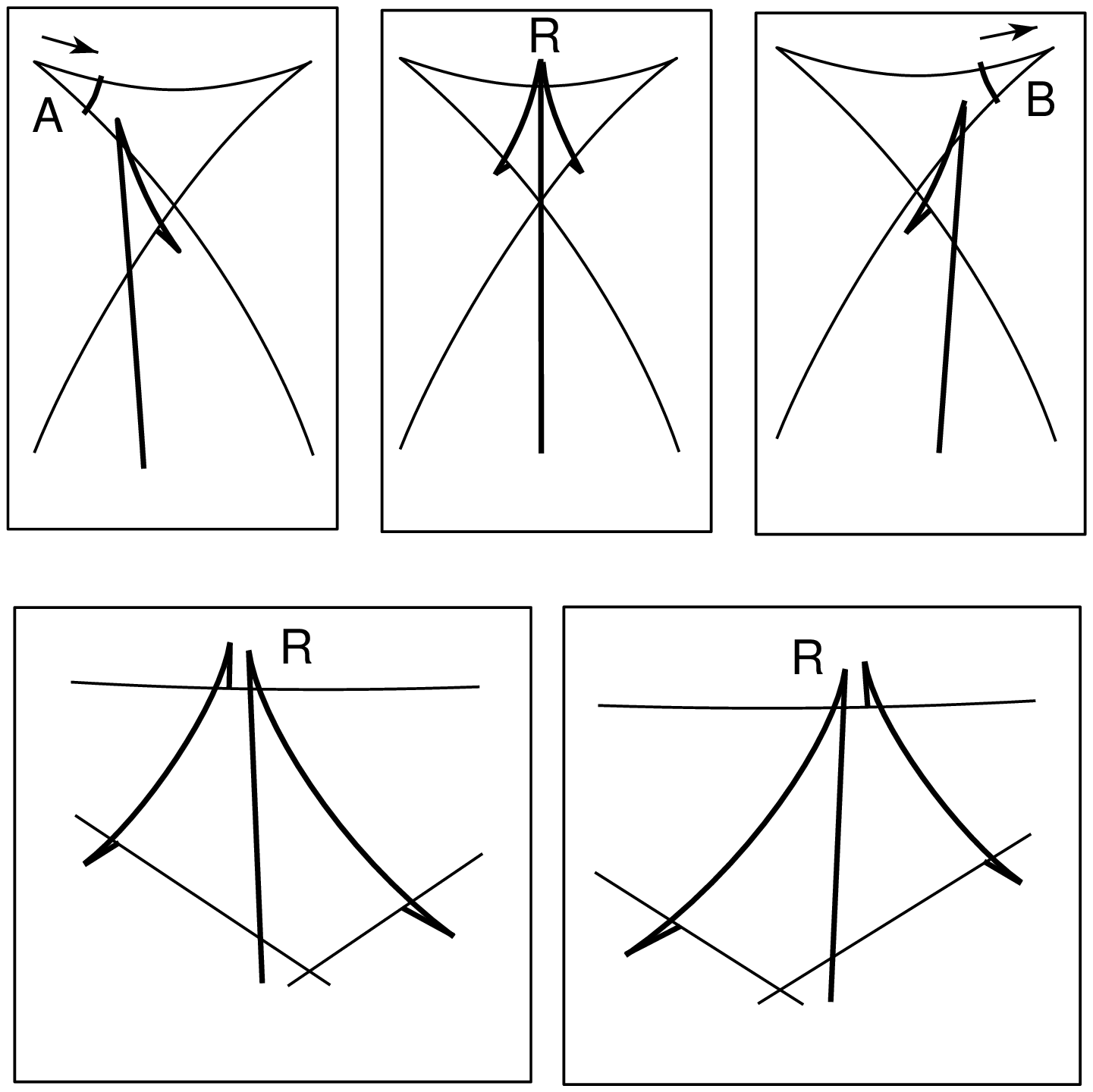}

\fignum Exotic solutions.
\end{center}
\end{figure}

\noindent$\bullet$ 
in point A new short  string appears from vacuum;

\noindent$\bullet$ 
in point R it recombinates with the long string: 
is attached to the long string while a part
of the long string is detached; 

\noindent$\bullet$ 
in point B short string disappears.

\vspace{1mm}
Therefore, these WS correspond to the processes of
creation, annihilation and recombination of strings. 
Further analysis \cite{exotic} shows 
that the density of energy for solutions of this kind
is not everywhere positive: the regions, marked $(+)$
on \fref{col-plate}g, possess positive energy, 
while the regions $(-)$ have negative energy
(theoretical physics uses for such solutions term ``exotic matter'' 
\cite{warp-drive}).
For such solutions the components of strings, which 
appear/disappear in vacuum near points A,B, 
have zero total energy-momentum and orbital momentum,
so that conservation laws do not prevent these processes.
It was also shown in \cite{exotic} that WS of this
kind are time-like, so that the square root 
in the Lagrangian of the string is real-valued,
however the cusp lines of WS correspond to branching points of 
this square root, therefore there are several 
choices of sign in Lagrangian. Exotic solutions correspond to such
a choice, that the areas of the parts marked $(\pm)$ on \fref{col-plate}g
give opposite contribution to the action. It was shown in \cite{exotic}
that exotic solutions are present at arbitrary dimension of the space-time 
and occupy regions, i.e. are not rare in the phase space of 
covariant Hamiltonian string theory. This phase space 
is formed by the coefficients of Fourier expansion
of the function $a(\s)$. Rejection of exotic solutions from 
the theory, possible only by means of explicit requirement 
$a_{0}(\s)>0$, or equivalent requirement in terms of Fourier coefficients, 
creates additional difficulties in quantum theory and actually is never done.

\section{Break of the string}

Considering string's breaking or any other transmutation process,
it is needed to fix the topological class of the process 
and find the extremum of action
in this class. For instance, breaking of open string into two open strings
corresponds to the diagram \fref{f2}, right. 
The initial and final positions of the strings should be fixed, 
and the surface should be varied to achieve the extremum of area,
assuming the boundaries and breaking point free.
Such surfaces can be reconstructed using the following algorithm
\cite{Artru}, see \fref{col-plate}i:

1. Let's represent the WS of open string by a supporting curve Q
(which is as usual 2P-periodical and light-like). Q is the 1st edge of the WS;
Q+P is the 2nd edge.

2. Take two arbitrary points A,C on the curve Q, lying inside one period.
Take their middle B = (A+C)/2. This will be {\it breaking point}
on the WS. Take also points A'=A-2P and B'=(A'+C)/2=B-P.

3. Consider the following curves:
AB, obtained from a segment AC of supporting curve, by homothetic 
contraction to the point A with coefficient 1/2;
BC, obtained from AC by homothetic contraction to the point C 
with the same coefficient (AB and BC are congruent);
and analogously:
A'B' is A'C 1/2-contracted to A';
B'C is A'C 1/2-contracted to C (A'B' and B'C are congruent).
Let A''BC' = A'B'C+P (parallel translation by P).

\vspace{2mm}\noindent
{\it{}Remark:} Curves ABC and A''BC' belong to the WS.
These curves mark a path of light signals emitted from
the point B on the WS and are called {\it characteristics}.

\vspace{1mm}
4. Consider triangular parts of the WS:

1 = part, restricted by arcs AB, BC and AC;

2 = part, restricted by arcs A''B, BC' and A''C'.

\noindent 
Shift these parts repeatedly by vectors BA and BC':

1' = 1+BA, 1'' = 1'+BA, ...

2' = 2+BC', 2'' = 2'+BC', ...

\noindent 
The sequence of images $\{1,1',1'',...\}$ forms a connected surface
(parts match each other along the characteristic AB and its images;
connection is continuous but generally not smooth --
the WS has a fracture along characteristics).
Construct $\{2,2',2'',...\}$ analogously. 

5. A part of initial WS, restricted by edges Q,Q+P and arcs A''B, BC
(lying on the left of A''BC), 
unified with the surfaces $\{1,1',1'',...\},\{2,2',2'',...\}$,
form a complete WS for decay ``open $\to$ 2 open''.

\vspace{2mm}\noindent
{\it{}Remarks:}

a) Products of decay $\{1',1'',...\},\{2',2'',...\}$ 
can be generated from the supporting curves
by the common rule ``locus of middles'', \fref{f4}. 
Supporting curves in this case are periodical continuations
of arcs CA and A''C'. Semi-periods are equal to energy-momentum,
and the property $P=P_{1}+P_{2}$ (conservation of energy-momentum 
in the decay process) is evident on \fref{col-plate}i.

\begin{figure}\label{f20}
\begin{center}
~\epsfysize=4cm\epsffile{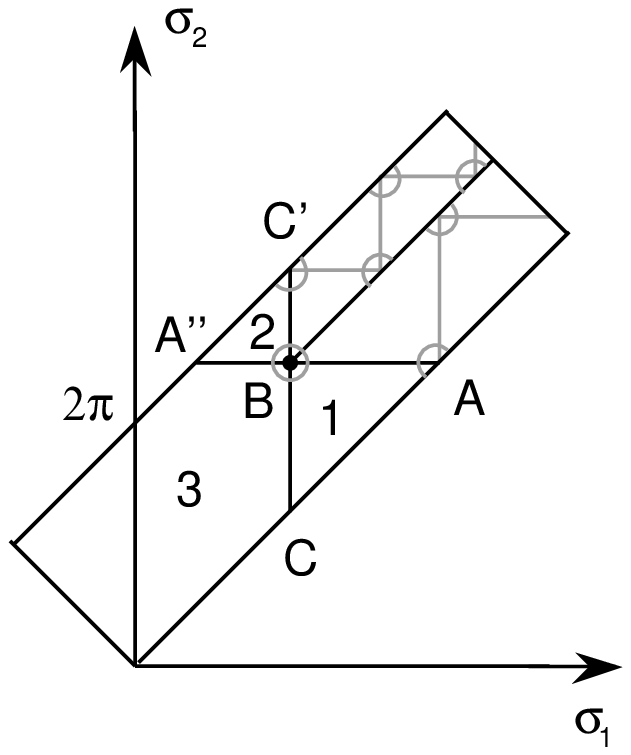}

\fignum Break ``open $\to$ 2 open'' on parameters plane.
\end{center}
\end{figure}

b) On the plane of parameters $(\s_{1},\s_{2})$ the characteristics
AB, BC, A''B, BC' correspond to straight lines $\s_{i}=Const$.
The parts 1,2 are triangles, restricted by these lines.
From here we have an equivalent algorithm of WS reconstruction: 
take triangles 1,2 on the parameters
plane, map them into space-time using 
$x(\s_{1},\s_{2})=(Q(\s_{1})+Q(\s_{2}))/2,$
obtain their images in translations, described above, 
and take the union of surfaces $\{1,1',1'',...\}\cup\{2,2',2'',...\}\cup3$ 
to represent the complete WS.

\vspace{2mm}\noindent
{\bf Theorem 9} (extremal property): 
WS, constructed by such algorithm, has extreme area.

\vspace{2mm}\noindent
{\bf Theorem 10} (break in a singular point): 
the products of string breaking have no fracture along the characteristics
if and only if B is singular point of original WS (see \fref{col-plate}h).

\def\figpage{ \newpage ~\vspace{-4mm}

\noindent
\begin{minipage}[t]{17cm}

\begin{figure}\label{col-plate}
\begin{center}
~\epsfxsize=15cm\epsfysize=15.9cm~\epsffile{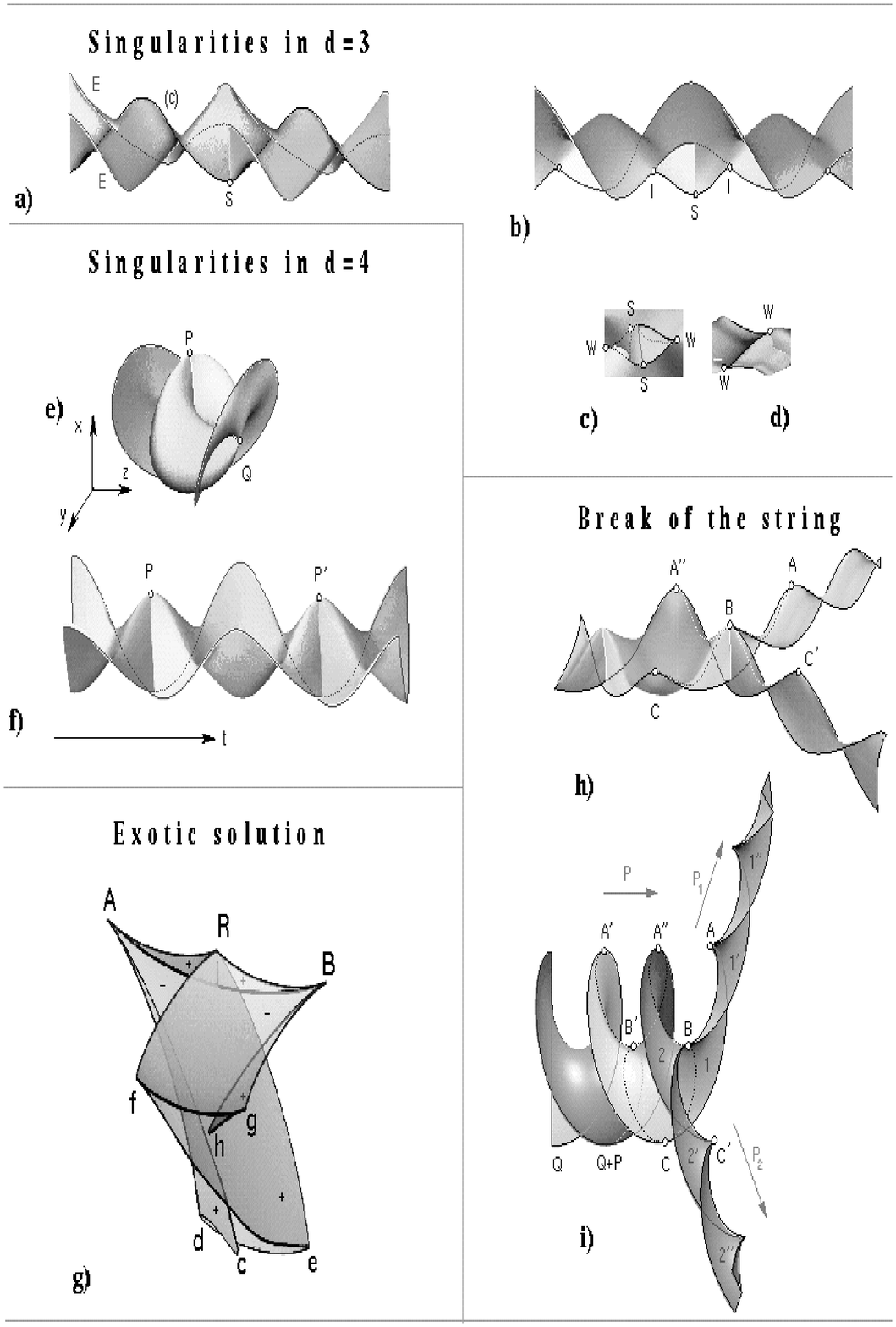}

\vspace{1mm}
\fignum Singular world sheets. 
Images are created by computer program \cite{sv},\\
representing string dynamics in Avango$^{\rm TM}$ Virtual Environment 
\cite{Avango}.

\end{center}
\end{figure}

\vspace{-2mm}~
\end{minipage} }
\twocolumn[\figpage]

\section*{Conclusion} 
We describe a geometrical method for explicit representation 
of solutions in string theory, convenient to study singular points on them. 
It is shown that in Minkowski space-time of dimension 3 and 4 
world sheets have stable singular points, which cannot be eliminated by small 
deformations of the surface in considered class. In dimension 3 
singularities are cusps, propagating along WS 
at light velocity, appearing/disappearing alone on the boundary 
of WS or by pairs in inner regions of WS. 
In dimension 4 singularities are isolated points, 
looking like instantaneous cusps periodically located on WS. 
Higher dimensional cases have no stable singularities. 

\vspace{1mm}
We show that at definite conditions the string theory has 
solutions of type $I\times{\bf R}^{1}$ whose boundary is not 
immersed in Minkowski space-time, but has cusp-like singularities. 
These solutions possess not everywhere positive density of energy 
and correspond to spontaneous creation of strings from vacuum.
Such processes lead to instability of vacuum state in string theory, 
mentioned earlier in work \cite{Zheltukhin}.

Considering the processes of string breaking by the scheme \cite{Artru}, 
we have found certain relations between singularities on strings
and the breaking processes. Particularly, smooth WS 
can appear only as a result of breaking of singular WS 
in one of the singular points. This fact gives
a possibility to construct models of decay of elementary particles,
where smooth WS describe the particles with long lifetime,
while singular WS decay finally to the smooth ones by a sequence of breaks
in the singular points. The dimensions $d=3,4$ are naturally selected
by these models as the only values of dimension where stable
singularities of WS exist. Consideration of these models on quantum level
is possible at least for subsets of phase space 
[3-5], 
admitting anomaly-free quantization at $d=3,4$.

\def\vsp{\vspace{-2mm}}

\subsection*{Appendix}

Here we present the proofs for the theorems stated above.

\vspace{2mm}
T1: Let $x(\s,\tau)$ be an extremal surface, 
presented in a neighborhood of a regular point $(\s_{0},\tau_{0})$
in conformal parametrization: $\dot x^{2}=-x'^{2}>0,\ \dot xx'=0$.
Consider a local variation $x\to x+\epsilon\delta x$, 
i.e. $\delta x\in C^{\infty}$ is vanishing outside of
a small vicinity of the point $(\s_{0},\tau_{0})$. 
Note that parametrization on disturbed surface 
is not conformal any more. Computing variation of the area:
\begin{eqnarray}
&&\delta((\dot xx')^{2}-\dot x^{2}x'^{2})^{1/2}=
\epsilon\;\delta\L_{1}+\epsilon^{2}\delta\L_{2}+O(\epsilon^{3}),\nn\\
&&\delta\L_{1}=\dot x\delta\dot x-x'\delta x',\  
\delta\L_{2}=(\dot x^{2}(\delta\dot x^{2}-\delta x'^{2})+\nn\\
&&((\dot x-x')(\delta x'-\delta\dot x))((\dot x+x')(\delta x'+\delta\dot x)))
/2\dot x^{2}.\nn
\end{eqnarray}
The terms linear in $\delta x$ give contribution
$\delta A_{1}=\epsilon\int \delta\L_{1}d\tau d\s =
\epsilon\int (-\ddot x+x'')\delta x d\tau d\s=0$
(here boundary terms vanish because $\delta x$ is local,
$\ddot x=x''$ is the condition of extremum in conformal
parametrization). Let's consider variations of a special form:
$\delta x(\s,\tau)=(0,0,F(\s,\tau),0...)$ in a system of coordinates
where $\dot x(\s_{0},\tau_{0})=(c,0,0...),\ x'(\s_{0},\tau_{0})=(0,c,0...)$,
i.e. the variation orthogonal to a tangent plane to WS in the point
$(\s_{0},\tau_{0})$. For such variations
$\delta\L_{2}=(F'^{2}-\dot F^{2})R/2,$ where
$R=1+(\dot x_{2}^{2}-x_{2}'^{2})/\dot x^{2}$.
Due to $\dot x^{2}=-x'^{2}>0,\ \dot xx'=0$ we have inequality $R\geq0$,
and because $R_{0}=1$ in the point $(\s_{0},\tau_{0})$,
we have $R>0$ in the vicinity of this point.
Now it's clear that $\delta\L_{2}$ is not positively defined,
however there is a possibility that $\delta A_{2}$ will be positive
after the integration. Let's consider an explicit example:
$F(\s,\tau)=f(\s^{2}/a^{2}+\tau^{2}/b^{2}),$ where $f(\rho)\in C^{\infty}$ 
is monotonous in $\rho\in[0,1]$ and $f(\rho)=0$ for $\rho>1$.
In the limit of small $a,b$ we have
$\delta A_{2}=\epsilon^{2}\int \delta\L_{2}d\tau d\s =
2\pi\epsilon^{2}ab(a^{-2}-b^{-2})I$, 
where $I=\int_{0}^{1}f'^{2}\rho^{3}d\rho>0$,
so that $\delta A_{2}<0$ for $a>b>0$ (maximum)
and $\delta A_{2}>0$ for $0<a<b$ (minimum), see \fref{f1}.

\vspace{2mm}
T2: the tangent plane to open WS on the edge is spanned on
two vectors $(Q',Q'')$ and is isotropic due to $Q'^{2}=Q'Q''=0$.
The tangent vector to the string $x'$, obtained as equal-time slice
of WS $x'_{0}=0$, is contained in the tangent plane to WS, and 
on the edge it is orthogonal to $Q'$: $x'Q'=0\ \Rightarrow\ \vec x'\vec Q'=0,$
so that the direction of the string at the end point 
and the velocity of the end point are orthogonal.

\vspace{2mm}
T3: The curves, which {\it are not} in general position,
define a closed stratified submanifold in the space of multijets
$_{2}J^{3}(S^{1},{\bf R}^{n})$ \cite{Mazer}, whose codimension
is greater than the dimension of considered mapping.
Due to criterions, given in \cite{Arnold-lectures},
this case can be eliminated by a small variation of the mapping.
The statement of the theorem follows from Thom's transversality 
theorem \cite{Thom,Arnold}, generalized to multijet space in \cite{Mazer}.

\vspace{2mm}
T4: The normal forms are given by the lowest order linear independent
terms in Taylor's expansion of WS near the singular point of each type.
The central part of the theorem is a proof that the higher order terms
do not change the structure of singularity and can be compensated by
LR-diffeomorphisms. For instance, Taylor's expansion of WS written 
for the case (c) in Rohrlich gauge in light coordinates, related with 
the direction $Q_{1}'=Q_{2}'$, contains together with the normal form 
$(x_{0},x_{2},x_{-})=(v,u^{2},u^{3})$ the additions to 
$x_{2},x_{-}$-components of the form $v^{k}u^{n}$ with 
the following set of possible indices 
$(k,n)\in\{k\geq2,n=0\}\cup\{k=0,n\geq3\}\cup\{k\geq1,n\geq2\}$,
where the term with $k=0,n=3$ enters to $x_{2}$-component only.
These terms can be presented as $f_{kn}=x_{0}^{k}x_{2}^{n/2}$ for even $n$
and $f_{kn}=x_{0}^{k}x_{-}x_{2}^{(n-3)/2}$ for odd $n$
(i.e. they belong to an ideal, spanned on monomials $v,u^{2},u^{3}$),
so that their addition is equivalent to smooth mapping
$x\to x+f_{kn}(x)$. All these mappings have unit Jacoby matrix at $x=0$
except of the case $k=0,n=3$, which corresponds to invertible 
linear mapping: $(x_{0},x_{2},x_{-})\to (x_{0},x_{2}+cx_{-},x_{-})$,
preserving the directions $(1,0,0)$ and $(0,1,0)$.
Convergent Taylor's series $\sum c_{kn}v^{k}u^{n}$
correspond to convergent $\sum c_{kn}f_{kn}(x)$,
so that the constructed mapping and its inverse are analytical.
Therefore, higher order corrections to the normal form (c)
can be compensated by analytical L-diffeomorphism, preserving light-like 
direction and related isotropic plane. In the case 
$(S):\ (v,u^{2},vu^{3})$ the terms $u^{2n+1}$
cannot be compensated by L-diffeomorphisms (they do not belong to
the ideal). The obstacle here is a line of self-intersection,
whose position on the parameters plane $v=0$ is invariant under
L-diffeomorphisms, while is changed by addition of $u^{2n+1}$. 
It's easy to prove that these terms can be removed 
by analytical R-diffeomorphisms (reparametrizations). In the same way
the transformation to normal form is performed
for other singularities. In each case it's needed to verify
that the constructed mapping is analytical and its
linear part is defined by non-degenerate Jacoby matrix of upper triangular
form: $J={\tiny 
\left(\begin{array}{ccc}*&*&*\\0&*&*\\0&0&*\end{array}\right)}$.
The last property ensures that the linear part of constructed
mapping preserves the light-like direction and isotropic plane,
initially related with $Q_{1}'=Q_{2}'$.

\vspace{2mm}
T5: if two hodographs on the circle $S^{1}$ have no intersection,
one of them should cover an arc with angular size $\Delta\ph<\pi$.
This is impossible due to condition $\oint d\s e^{i\ph}=0$,
equivalent to closeness of the curve $\vec Q(\s)$.

\vspace{2mm}
T6: in Rohrlich gauge the topological charge of cusp 
is defined by orientation of a pair $(\vec Q'_{1},\vec Q''_{1}+\vec Q''_{2})$
and is equal to $\sign(\ph'_{1}+\ph'_{2})$. This number is a particular case
of the following topological invariant. Let's consider a mapping
$S^{1}\times S^{1}\to S^{1}$, defined by the function 
$f(\s_{1},\s_{2})=-\ph_{1}(\s_{1})+\ph_{2}(\s_{2})$.
The cusp line (c) is zero level $f(\s_{1},\s_{2})=0$.
The vector $(-\ph'_{1},\ph'_{2})$ is a normal
and $(\ph'_{2},\ph'_{1})$ is tangent element to (c),
which can be used to define the orientation of (c). 
Let's consider another oriented contour (k) on the torus,
with tangent element $(d\s_{1},d\s_{2})$, intersecting (c) 
in a certain point. The number $\nu=\sign(-\ph'_{1}d\s_{1}+\ph'_{2}d\s_{2})$ 
in this point is called {\it index of intersection}.
Let's also consider a mapping $S^{1}\to S^{1}$, defined
by a restriction of $f(\s_{1},\s_{2})$ to the contour (k).
The same number $\nu=\sign(df)$, estimated in pre-images 
of point $0$ for this mapping (i.e. in (c)$\cap$(k)),
is called {\it degree of mapping}.
According to the theorems, proved in \cite{DNF}, the sum of such numbers 
estimated in all points of intersection (c)$\cap$(k),
is invariant under homotopies of mapping $f$ and contour (k).
In our case it equals $-n_{1}(k)+n_{2}(k)$, where $n_{i}(k)$
are numbers of revolutions of $\ph_{i}(\s_{i})$ corresponding to
a complete passage of (k). The statement of the theorem corresponds to
the case when (k) is equal-time slice:
$(\s_{1},\s_{2})=(\tau_{0}-\s,\tau_{0}+\s),\ \s\in[0,2\pi]$,
so that $\nu=\sign(\ph'_{1}+\ph'_{2})$ and $\sum\nu_{i}=n_{1}+n_{2}$, 
where are $n_{i}$ are numbers of revolutions of $\ph_{i}$
for basis cycles on torus. Another invariant $n_{1}-n_{2}$
equals to the intersection index of (c) with the trajectory of
point $\s_{0}$ on the string: $(\s_{1},\s_{2})=(\tau-\s_{0},\tau+\s_{0}),\ 
\tau\in[0,2\pi]$, representing the total number of windings
of cusp lines around the cylinder of closed string WS.

\vspace{2mm}
T7: if the functions $\ph_{i}$ are monotonous and 
$\sign\ph'_{1}=\sign\ph'_{2}$, then on equal-time slice
we have a function $f(\s)=-\ph_{1}(\tau_{0}-\s)+\ph_{2}(\tau_{0}+\s)$,
which is also monotonous and has the interval of variation
$f(2\pi)-f(0)=2\pi(n_{1}+n_{2})$. In this case the equation 
$f(\s)=2\pi k,\ k\in{\bf Z}$ has on the interval $\s\in[0,2\pi)$
exactly $|n_{1}+n_{2}|$ isolated solutions. Other statements
of the theorem follow from T6.

\vspace{2mm}
T8: intersection of cusp lines corresponds to
a saddle point of the function
$f(\s_{1},\s_{2})=-\ph_{1}(\s_{1})+\ph_{2}(\s_{2})$
on the level $f=0$. Small variations transform
the cusp lines similarly to recombination of hyperbolas $x^{2}-y^{2}=c$, 
when $c$ passes through $0$. Then, using conditions 
$\ph'_{1}+\ph'_{2}=0,\ d(\ph'_{1}+\ph'_{2})=
(\ph''_{1}-\ph''_{2})d\s_{1}\neq0,\ d\s_{2}=-d\s_{1}$,
valid in the point W, we see that the function $\ph'_{1}+\ph'_{2}$
changes sign in the passage through this point, so that
created cusps have opposite topological charges,
while the velocities $\vec Q_{1,2}'$ are equal.

\vspace{2mm}
T9: internal regions of the patches
$\{1,1',1'',...\},$ $\{2,2',2'',...\},3$ 
are constructed by the rule $x(\s_{1},\s_{2})=(Q(\s_{1})+Q(\s_{2}))/2,$
so that Lagrange-Euler equations $\df_{i}p_{i}=0$ are satisfied there. 
It is only needed to check that condition 
$\Delta I=\Delta p_{i}\epsilon_{ij}d\s_{j}=0$ 
is satisfied on the lines of connection (characteristics),
so that the flow of momentum, coming out one patch,
is equal to the flow of momentum, coming into another patch.
Here $\Delta p_{i}$ is a discontinuity of momentum
on the characteristics. Computing $p_{i}$:
$p_{1}=Q'(\s_{2}),\ p_{2}=Q'(\s_{1})$ (see \cite{exotic}), we have
$\Delta I=-\Delta Q'(\s_{1})d\s_{1}+\Delta Q'(\s_{2})d\s_{2}$,
where $\Delta Q'(\s_{i})$ are discontinuities of tangent vector
to supporting curve. Then, because the discontinuity 
$\Delta Q'(\s_{i})\neq0$ propagates along the characteristic 
$d\s_{i}=0$ with the same $i$, we have 
$\Delta I=0$. Actually, a discontinuity occurs only for 
components of momentum, tangential to characteristics.
The flow of momentum through the free edge vanishes
due to the identities $Q'(\s_{1})=Q'(\s_{2}),\ d\s_{1}=d\s_{2}$.
Then, considering the contours, separating 
the break point and the points (A,C'...), where characteristics 
intersect the edges, \fref{f20},
we see that the flow of momentum through these contours
is not changed when the contour is continuously deformed 
to a point. Then, using the fact that $p_{i}$ are bounded:
$|p_{i}|<Const$ (even in the singular point of WS),
we see that flow of momentum vanishes in this limit, 
therefore the momentum has no leakage in these points.

\vspace{2mm}
T10: string breaks in singular point, when the 
tangent vectors to supporting curve in points A,C 
are parallel. In this case the periodical continuation 
of curves CA and A''C' on \fref{col-plate}h is $C^{1}$-smooth.

\end{document}